      [1999/12/01 v1.4c Il Nuovo Cimento]
\documentclass{cimento}
\usepackage{graphicx}  
\usepackage{psfig}

\newcommand{\ApJL}{Astrophys. J. Lett.}
\newcommand{\ApJ}{Astrophys. J.} 

\newcommand{\PRD}{Phys. Rev. D}
\newcommand{\MNRAS}{Mon. Not. Roy. Astron. Soc.}

\def\sun{\hbox{$\odot$}}

\newcommand{\da}{d_A}

\newcommand{\dop}{{\rm dop}}
\newcommand{\ksz}{{\rm Kin SZ}}
\newcommand{\bn}{\hat{\bf n}}    
\newcommand{\bk}{\hat{\bf k}}

\newcommand{\bfk}{{\mathbf{k}}}

\newcommand{\vecx}{{\bf x}}
\newcommand{\veck}{{\bf k}}

\newcommand{\vecl}{{\bf l}}

\newcommand{\rad}{r}

\newcommand{\bfl}{{\mathbf{l}}}

\newlength{\tskip}
\newlength{\colwidth}

\newcommand{\sz}{{\rm SZ}}
\newcommand{\bp}{{\cal C}}

\newcommand{\beq}{\begin{equation}}
\newcommand{\eeq}{\end{equation}}
\newcommand{\beqa}{\begin{eqnarray}}
\newcommand{\eeqa}{\end{eqnarray}}

\def\VEV#1{{\langle #1 \rangle}}
\long\def\comment#1{}

\title{Statistical Imprints of SZ Effects \\
in the Cosmic Microwave Background}
\author{Asantha Cooray\from{ins:cal}\from{ins:uci}\thanks{asante@caltech.edu}, Daniel Baumann\from{ins:pri}, Kris Sigurdson\from{ins:cal}}
\instlist{\inst{ins:cal} Theoretical Astrophysics, California Institute of Technology, Pasadena, CA 91125, USA
\inst{ins:uci} Department of Physics and Astronomy, 4129  Frederick Reines Hall, University of California,   Irvine, CA 92697, USA
\inst{ins:pri} Department of Physics, Jadwin Hall, Princeton University, Princeton, NJ 08544, USA}
\begin{document}

\maketitle

\begin{abstract}
We review several aspects of the Sunyaev-Zel'dovich (SZ) effect associated with the large scale baryon distribution and its characteristic signatures in the statistics of cosmic microwave background (CMB) anisotropies.
We discuss (1) the contributions to the angular power spectrum from the thermal and kinetic SZ effects,
(2) the effect of SZ non-Gaussianities on cosmological parameter estimation, and (3)
the SZ-induced CMB polarization towards galaxy clusters. We also discuss the extent to which recent measurements of small angular scale CMB anisotropy can be accounted for by SZ clusters and the potential use of SZ polarization towards large samples of galaxy clusters as a probe of dark energy parameters.
\end{abstract}

\tableofcontents

\newpage

\section{Introduction}

The Sunyaev-Zel'dovich effect (SZ; \cite{SunZel80}) is the upscattering of cosmic microwave background (CMB) photons as they pass through the hot gas of a galaxy cluster.  This upscattering introduces frequency-dependent secondary temperature anisotropies in the CMB that are proportional to the integral of the
electron pressure $\Pi_e = n_e k_B T_e$ along  a given line of sight through the Universe:
\begin{equation}
\frac{\Delta T^{\rm SZ}}{T_{\rm CMB}}\equiv g_{\nu}(x) y=g_{\nu}(x)\int \, d\eta a(\eta) \frac{\sigma_{\rm T}}{m_{\rm e}}\Pi_e(\eta)  \, .
\label{eqn:sz}
\end{equation}
where we have defined $y$, the Compton $y$-parameter, $\sigma_{\rm T}$ is the Thomson cross-section, and $m_{\rm e}$ is  the electron rest-mass \footnote{Note that here and in what follows we adopt units where the speed of light $c$ is set to unity.}. The frequency dependence of the SZ anisotropies is accounted for by the function $g_{\nu}(x)=x{\rm coth}(\frac{x}{2})-4$ with $x={h\nu}{(k_B T_{\rm CMB})^{-1}}$, which, in the low frequency Rayleigh-Jeans part of the spectrum has the limit $g_{\nu} \rightarrow -2$.  This form for the frequency spectrum leads to an apparent temperature decrement, relative to the primordial CMB black body spectrum, at frequencies below the null at $x_{\rm cross}$, and an increment above  $x_{\rm cross}$.  This SZ null occurs at approximately $217~{\rm GHz}$ with a slight dependence on the electron temperature $T_{e}$ due to higher-order relativistic corrections to the inverse-Compton scattering process neglected in the Thomson approximation.  In Eq.~(\ref{eqn:sz}), $\eta \equiv \eta(z)$ is the conformal distance 
from the observer at redshift $z=0$ to the cluster at redshift $z$, given by
\begin{equation}
\eta(z) = \int_0^z {dz' \over H(z')} \, ,
\end{equation}
where the expansion rate of cold dark matter (CDM) cosmological models with a cosmological constant $\Lambda$ is
\begin{equation}
H^2 = H_0^2 \left[ \Omega_m(1+z)^3 + \Omega_K (1+z)^2
              +\Omega_\Lambda \right]\, .
\end{equation}
The Hubble distance today is $H_0^{-1} = 2997.9 \, h^{-1} $Mpc with $h =0.72 \pm 0.08$~\cite{Freedman:2000cf, Spergel:2003cb}. $\Omega_i$ denotes the contribution of component $i$
 in units of the critical density $\rho_{\rm crit} \equiv 3H_0^2/8\pi G$, with
$i=c$ for the CDM, $b$ for the baryons, and $\Lambda$ for the cosmological
constant. We also define the
auxiliary quantities $\Omega_m \equiv \Omega_c+\Omega_b$ and
$\Omega_K \equiv 1-\sum_i \Omega_i$, which represent the total matter density and
the contribution of spatial curvature to the expansion rate,
respectively.\\

The thermal SZ effect has been imaged in the CMB towards massive galaxy clusters whose presence is a priori known from optical data \cite{Caretal96,Jonetal93}. 
The temperature of the scattering medium in certain clusters can reach up to 10 keV producing temperature changes in the CMB of order 1 mK at Rayleigh-Jeans (RJ) wavelengths. The individual galaxy cluster SZ images have a variety of astrophysical and cosmological applications
including a direct measurement of the angular diameter distance to the cluster through a combined analysis with X-ray data and
a measurement of the gas mass, and, thus the baryon fraction of the universe. We do not discuss these applications of the SZ effect in the present review, but
rather will concentrate on the statistical study of wide-field CMB data where SZ effects lead to anisotropies in the temperature
distribution both due to resolved and unresolved galaxy clusters. In fact, the thermal SZ contribution is the
dominant signal beyond the damping tail of the primary anisotropy power spectrum.\\

In addition to the thermal SZ effect, the bulk flow of electrons in galaxy clusters and other virialized halos 
that scatter the CMB photons leads to a second contribution to temperature fluctuations, the kinetic SZ effect. This effect arises from the well known Doppler effect 
\cite{Kai84,CooHu00}:
\begin{equation}\label{equ:dop}
\frac{ \Delta T^\dop}{T_{\rm CMB}}(\bn) = \int_0^{\eta_0} d\eta g(\eta) \bn \cdot {\bf v}_g(\eta,\bn
\eta)\, ,
\end{equation}
where ${\bf v}_g$ is the baryon velocity and $g(\eta)$ is the visibility function (also interpreted as the
probability of scattering within $d\eta$ of $\eta$)
\begin{equation}\label{equ:visibility}
g =  \dot \tau e^{-\tau} = X_e(z) H_0 \tau_H (1+z)^2 e^{-\tau}\,.
\end{equation}
Here
$\tau(\eta) = \int_0^{\eta} d\eta' \dot\tau(\eta')$ is the optical depth out to distance $\eta$,
$X_e(z)$ is the ionization fraction as a function of redshift, and
\begin{equation}
       \tau_H = 0.0691 (1-Y_p)\Omega_b h\,,
\end{equation}
is the optical depth due to Thomson
scattering to the Hubble distance today, assuming full
hydrogen ionization with primordial helium fraction of $Y_p (=0.24)$.
From equation (\ref{equ:dop}) follows the secondary temperature anisotropy due to the kinetic SZ effect. The kinetic SZ fluctuation is written as an integral over the product of the radial velocity, $\bn \cdot {\bf v}_g$ , and the density fluctuation associated with the cluster, $\delta_g$: 
\begin{equation}\label{equ:kSZ}
\frac{\Delta T^\ksz}{T_{\rm CMB}}(\bn) = \int_0^{\eta_0} d\eta g(\eta) \bn \cdot {\bf v}_g(\eta,\bn
\eta) \cdot \delta_g(\eta, \hat{\bf n} \eta)\, .
\end{equation}

The angular power spectrum of anisotropies generated by the
Doppler effect peaks at angular scales
corresponding to the horizon at the time of scattering. The effect cancels out significantly at scales smaller than the horizon since photons scatter against the crests and troughs of the perturbation.
The kinetic SZ effect arises from the second-order modulation of the Doppler effect by non-linear density 
fluctuations associated with virialized objects such as galaxy clusters, and avoids the strong cancellation associated with the linear Doppler effect.  
The kinetic SZ effect is also known as the Ostriker-Vishniac effect \cite{OstVis86} when the density fluctuations are associated with the
 linear density field of the large-scale structure. In addition to modulation of the velocity field due to density perturbations,
the velocity field is also modulated by fluctuations in the ionization fraction of electrons in a partly reionized universe during the reionization epoch. This latter contribution is generally referred to as
patchy-reionization \cite{Aghanim:1996ib,Gruzinov:1998un,Santos:2003jb}. 
Due to the density weighting, the kinetic SZ effect peaks at small angular scales (sub arcminutes for $\Lambda$CDM).
For a fully ionized universe, contributions are broadly distributed in redshift so that the
power spectra are moderately dependent on the optical depth $\tau$.\\

In this review we focus on the statistical characterization of secondary CMB temperature fluctuations in the
form of the angular power spectrum of SZ-induced anisotropies. SZ maps have important additional applications including number counts of halos where a comparison to analytical predictions can be used to derive cosmological parameters. These studies will not be the subject of this short review, but are discussed elsewhere in these proceedings.\\

The outline of this article is as follows. In \S2 we derive the power spectrum of the thermal SZ effect using the halo model of large scale structure. \S3 compares these theoretical predictions to the current CMB data. In \S4 we derive the power spectrum of the kinetic SZ effect. Finally, we discuss secondary CMB polarization due to free electrons in galaxy clusters and its cosmological use as a powerful probe of dark energy in \S5.

\section{Thermal SZ Contribution to the CMB Power Spectrum}

This section reviews the thermal SZ contribution to the angular power spectrum of CMB anisotropies following standard derivations published in the literature \cite{ColeKai88,Coo00,Coo01,KomSel02b}.

The basic idea of the halo approach to large scale structure is summarized in Fig.~1.
The complex dark matter distribution is replaced by a population of dark matter halos that
reproduces the relevant statistical properties of the actual distribution \cite{CooShe02}.
The SZ anisotropy calculation is particularly well suited for the halo approach since high resolution simulations of the
hydrodynamics of clusters, including radiative cooling and feedback from supernovae and galactic winds,
indicate that nearly all of the power at small scales originates from regions associated with virialized halos rather
than from filamentary structures or the diffuse gas between halos \cite{White02}.  The semi-analytic calculations of
Refs.~\cite{KomSel02b,KomSel02a} are extended here by accounting for the self-gravity of the cluster gas and allowing for
scatter in the concentration-mass distribution.

\begin{figure}[t]
\centerline{\psfig{file=cooray1.eps,width=5.2in,angle=0}}
\caption{The complex distribution of dark matter (a) found in numerical
simulations may be  replaced with a distribution of dark
matter halos (b) with the mass function matching that found in
simulations.}
\label{fig:halos}
\end{figure}

To calculate the angular power spectrum associated with the thermal SZ effect, it is convenient to describe statistics in terms of the baryon gas pressure $\Pi_g$, which is related to the pressure of the electrons in the cluster by $\Pi_e(r) = \frac{2X+2}{3X+5}\Pi_g(r)$, where $X=0.76$ is the primordial hydrogen abundance. We consider the difference in the local gas pressure relative to the mean pressure
averaged over all positions, at a given epoch, and write the fluctuation as $\delta_\Pi (\vecx, t) \equiv (\Pi - \bar{\Pi})/\bar{\Pi}$, where $\Pi$ denotes $\Pi_e$ or $\Pi_g$. The mean value of $\delta_\Pi$ vanishes by definition:
\begin{equation}
\langle \delta_\Pi \rangle = \lim_{R \rightarrow \infty} \frac{\int_{|\vecx| < R} d^3\vecx 
\, \delta_\Pi(\vecx)}{4/3 \pi R^3} = 0 \, .
\end{equation}
The correlation function of pressure fluctuations is defined as $\xi(x) \equiv \langle \delta_\Pi(\vecx_1)\delta_\Pi(\vecx_2)\rangle$, where homogeneity and isotropy imply that correlations only
depend on the absolute separation of the two locations, $x \equiv |\vecx_1-\vecx_2|$. If the fluctuations are Gaussian, i.e. the random
field related to $\delta_\Pi$ has a Gaussian probability distribution, then
the distribution function is fully specified by the correlation function alone. A generic prediction of inflationary cosmology is the Gaussianity of the spectrum of initial fluctuations. However, non-linear growth and the formation of
virialized halos lead to significant non-Gaussianity in the pressure field at the low redshifts probed by the SZ effect.
Hence, to fully specify the statistics of the SZ power spectrum one must
make measurements beyond the correlation function.

Instead of the real space correlation function, we will discuss statistics primarily in Fourier space where the fluctuation in pressure is
\begin{equation}
\delta_\Pi(\veck) = \int d^3 \vecx \, \delta_\Pi(\vecx) e^{-i \veck \cdot \vecx} \, .
\end{equation}
It is easily shown that Gaussianity of $\delta_{\Pi}(\vecx)$ implies Gaussianity of $\delta_{\Pi}(\veck)$.
We
define the three-dimensional power spectrum of pressure fluctuations in the large scale structure as
\begin{equation}\label{equ:Press}
\langle \delta_\Pi(\veck) \delta_\Pi(\veck') \rangle \equiv (2\pi)^3\delta_D(\veck-\veck')P_{\rm \Pi \Pi}(k) \, .
\end{equation}
The Dirac delta function $\delta_D$ expresses the fact that wave modes
$\veck$ and $\veck'$ are independent and uncorrelated since the different $\delta_\Pi(\veck)$
have random phases.
Spatial isotropy requires the power spectrum to be independent of the phase of the wave vector and depend only on the magnitude of the vector.
The power spectrum and the correlation function are related via
\begin{equation}
P_{\Pi \Pi}(k) = \int d^3 \vecx \, \xi(x) e^{-i \veck \cdot \vecx} = \int dx \, 4 \pi x^2 \xi(x) \frac{\sin(k x)}{k x} \, .
\end{equation}

The halo model for large scale structure predicts two separate contributions to the power spectrum:
\begin{equation}
P_{\rm \Pi \Pi}(k;z)=P_{\rm \Pi \Pi}^{\rm 1h}(k;z)+P_{\rm \Pi \Pi}^{\rm 2h}(k;z) \, ,
\end{equation}
where
\begin{eqnarray}
P_{\rm \Pi \Pi}^{\rm 1h}(k;z) &=& I_{2,\rm \Pi \Pi}^{0}(k;z), \; \quad \; {\rm and} \;  \nonumber \\
P_{\rm \Pi \Pi}^{\rm 2h}(k;z) &=& [I_{1,\rm \Pi \Pi}^{1}(k;z)]^{2}P(k;z),
\label{eqn:phalo}
\end{eqnarray}
denote the power spectrum for correlations of gas pressure within a single halo (1h) and between two separate
halos (2h). It is found that the 1-halo term dominates on small angular scales. The functions $I_{2,\rm \Pi \Pi}^{0}$ and $I_{1,\rm \Pi \Pi}^{1}$
are defined below and $P(k;z)$ is the linear matter power spectrum as a function of redshift $z$.
Useful fitting formulae for the transfer function relating $P(k;z)$ to the primordial power spectrum $P(k)$ may be found in Ref.~\cite{EisHu99}.

\begin{figure}[t]
\centerline{\psfig{file=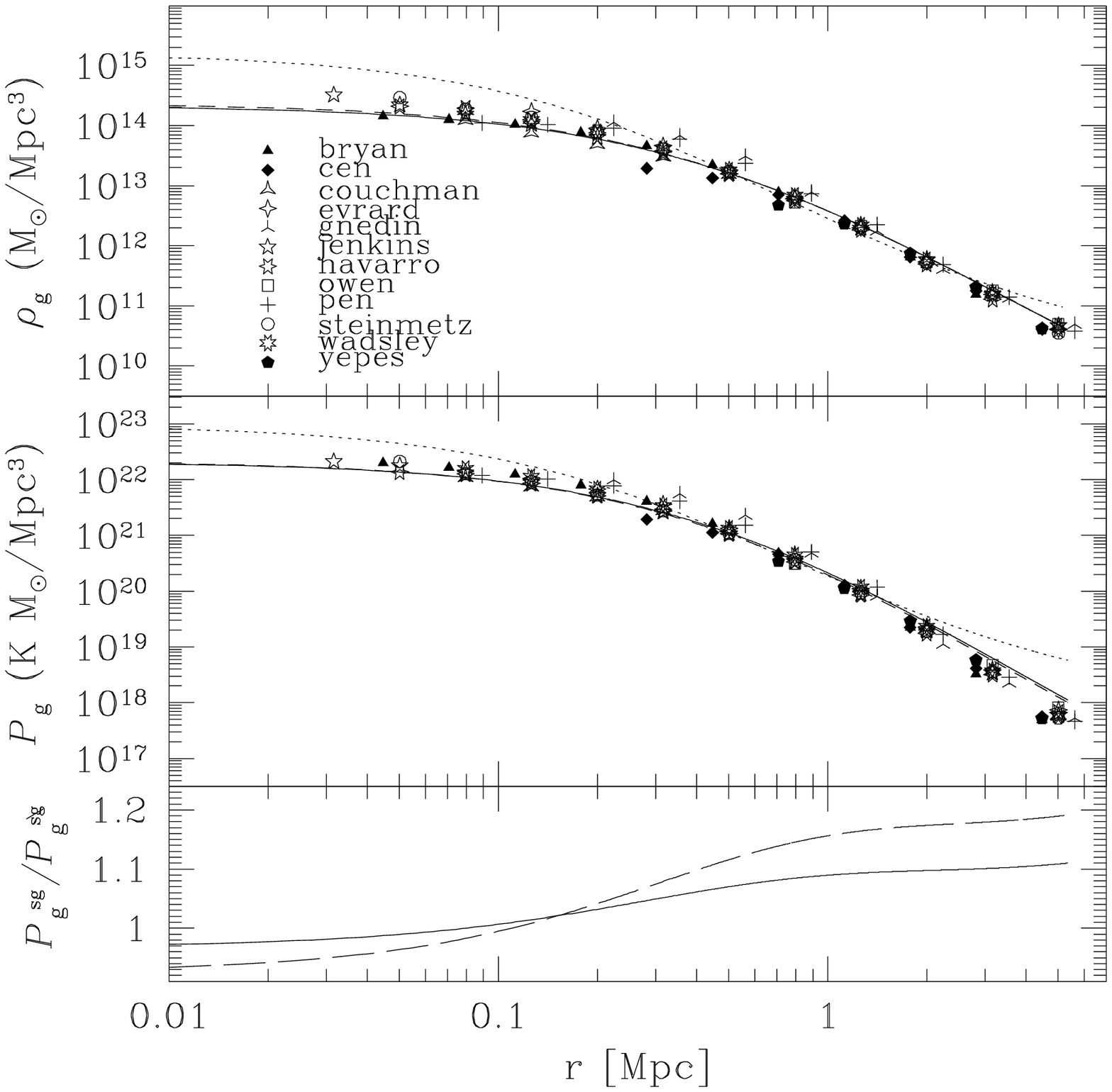,width=3.3in,angle=0}
\psfig{file=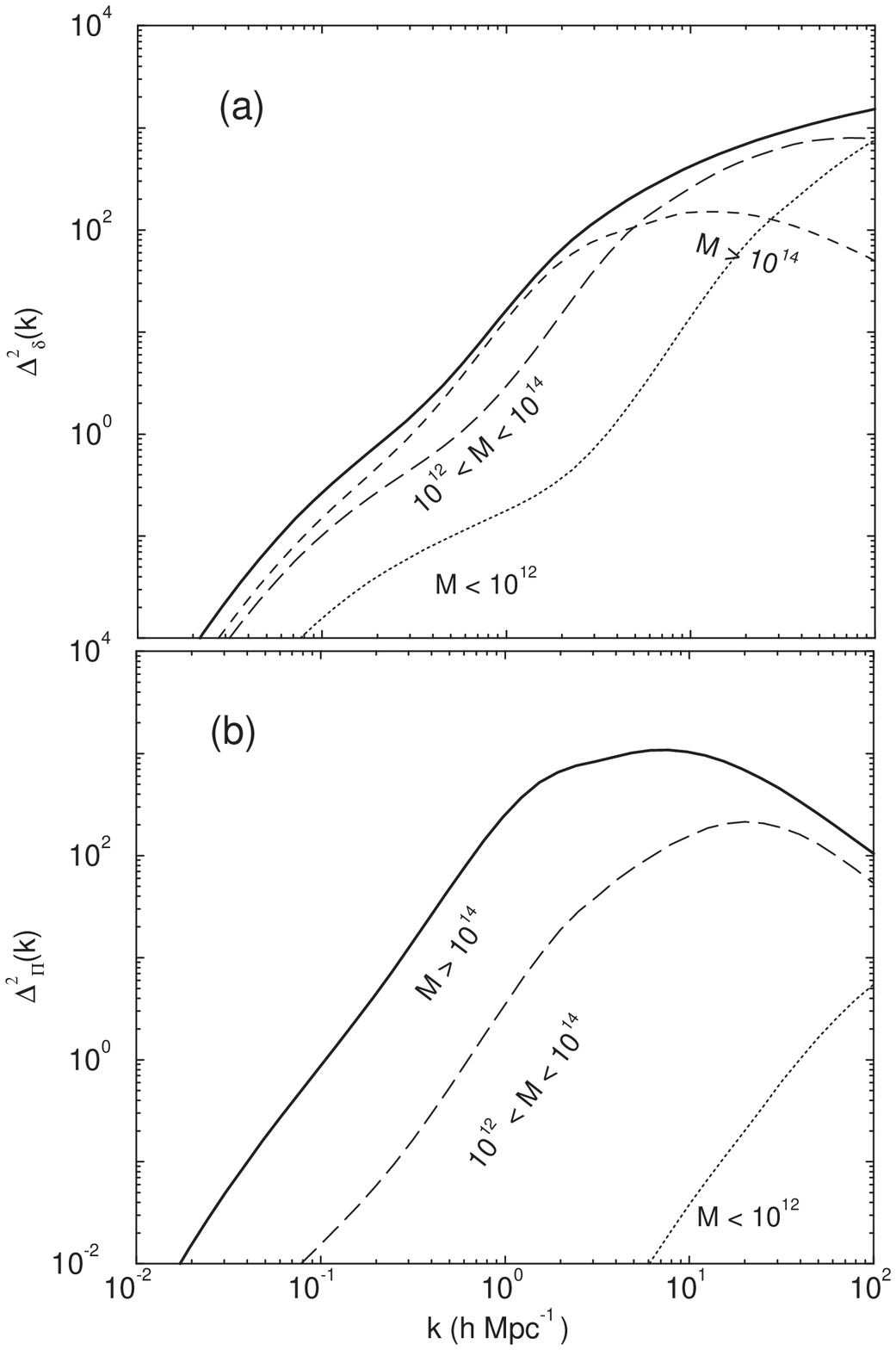,width=2.4in}}
\caption{{\it Left:} The density ({\rm top}) and pressure ({\rm middle}) profiles for the self-consistent self-gravitational model (solid) and the isothermal cluster
model (dotted) compared with numerical simulations from the Santa Barbara cluster comparison project.  The isothermal model overestimates the pressure and density in the inner and outer regions of the cluster.  The lower panel shows the ratio of the pressure profile calculated with self-gravity to the profile calculated without self-gravity for $f_{\rm b}=1/10$ (solid), the value used in the Santa Barbara cluster comparison project, and $f_{\rm b}=1/6$ (long-dashed), our fiducial value.  Self-gravity raises the pressure by $\approx 20 \%$ outside the core for our fiducial value. {\it Right:}
The mass dependence on the dark matter power spectrum (a) and the
pressure power spectrum (b). Here, we show the total contribution
broken up into mass limits as labeled in the figure.
As shown in (a), the large scale contribution to the dark matter power
comes from massive halos while small mass halos contribute at small
scales. For the pressure, in (b), only massive halos above a mass of
$10^{14}$ M$_{\sun}$ contribute to the power.}
\label{fig:SB}
\end{figure}
Using the three-dimensional Fourier transform of the (spherically symmetric) halo pressure profile $\Pi(r)$,
\begin{equation}\label{equ:pressFT}
\Pi_e(k)= \int d^3 {\bf r} \, \Pi_e(r) e^{-i \veck \cdot \vecx} =\int_0^{\infty} dr \, 4\pi r^2 \Pi_e(r) \frac{{\rm sin}(kr)}{kr} \, ,
\end{equation}
we can write the two $I$-terms in Eq.~(\ref{eqn:phalo}) as
\begin{eqnarray}
\label{eqn:iterms}
I_{2,\rm \Pi \Pi}^{0}(k; z) &\equiv& \int  dM \int dc \,\frac{d^2n}{dMdc} \left|\Pi_e(k, M, c ; z) \right|^2 \; \quad {\rm and}  \nonumber \\
I_{1,\rm \Pi \Pi}^{1}(k; z) &\equiv& \int  dM \int dc \,\frac{d^2n}{dMdc}b_{1}(M;z) \Pi_e(k, M, c ; z ) \, .
\end{eqnarray}
Here, ${d^2n}/{dMdc}$ is the bivariate halo mass function, the co-moving number density $n$ of halos between mass $M$ and $M+dM$ with concentration parameter
between $c$ and $c+dc$.  The concentration parameter $c = r_{\rm v}/r_{\rm s}$ is defined as the ratio of the virial radius of the cluster $r_{\rm v}$ and a characteristic scale radius $r_{\rm s}$.
In writing Eq.~(\ref{eqn:phalo}), we assumed that halos trace the linear density field such that the
halo power spectrum is $P_{\rm hh}(k;z|M) = b^2_1(M)P(k;z)$ where $M$ denotes the halo mass
 and the halo bias $b_1(M;z)$ is \cite{Moetal97}
\begin{equation}
b_{1}(M;z)=1+\frac{a\nu^2(M;z)-1}{\delta_c(z)} + \frac{2p}{\delta_c(z)\{1+[a\nu^2(M;z)]^{p}\}} \, ,
\end{equation}
where $(a,p)$ are numerical parameters and
$\nu(M;z) \equiv \delta_{\rm c}(z)/\sigma(M;z)$ as discussed in more detail in section 2.1.
We see from Eq.~(\ref{eqn:iterms}) that the two ingredients involved in calculating the power spectrum with the halo model
are the radial pressure profile of the gas within each halo and the mass function, which specifies the number of such halos.\\

The angular power spectrum of the thermal SZ effect is just proportional to a projection of the
three dimensional pressure power spectrum integrated along the line of sight.  To derive the angular spectrum, we take
 spherical harmonics of Eq.~(1) such that
\begin{equation}
\frac{\Delta T}{T}(\bn) = \sum_{lm} \left(\frac{\Delta T}{T}\right)_{lm} Y_{lm}(\bn) \, .
\end{equation}
Making use of the Rayleigh expansion of a plane wave
\begin{equation}
e^{i{\bf k}\cdot \hat{\bf n}\eta}=4 \pi \sum_{lm} i^l j_l(k\eta) Y_{lm}^\ast(\bk) Y_{lm}(\bn)\, ,
\label{eqn:Rayleigh}
\end{equation}
we find the spherical harmonic moments of the SZ map
\begin{equation}\label{equ:DT}
\left(\frac{\Delta T}{T}\right)_{lm} =   g_{\nu} \frac{\sigma_T}{m_e} \, \, 4 \pi i^l\int \frac{d^3\veck}{(2\pi)^3}\, \Pi_e(\veck) Y_{lm}^\ast(\bk) \int d\eta \,  a^{-2}(\eta) j_l(k \eta)  \, .
\end{equation}

The angular power spectrum of the SZ map, is
defined in terms of multipole moments $[\Delta T/T]_{lm}$ as
\begin{equation}
\langle [\Delta T/T]_{lm} [\Delta T/T]_{l' m'}^\ast \rangle = C_{l}^{\sz} \delta_{ll'} \delta_{m m'}  \, .
\end{equation}
Using equation (\ref{equ:DT}) and the relation
$\langle \Pi_e(\veck) \Pi_e(\veck') \rangle = \bar{\Pi}_e^2 \left[ \langle \delta_\Pi(\veck) \delta_\Pi(\veck') \rangle +1 \right]$ we obtain
\begin{eqnarray}
C_l^{\sz} = && g_{\nu}^2 \left(\frac{\sigma_T}{m_e}\right)^2 [\bar{n}_e(0)]^2 \, \frac{2}{\pi}\int k^2 dk P_{\Pi \Pi} (k) \times \nonumber \\
&& \times \int d\eta_1 \int d\eta_2  \, a^{-2}(\eta_1) a^{-2}(\eta_2)
j_l(k\eta_1) j_l(k\eta_2) \, ,
\end{eqnarray}
where $\bar{n}_e(0)$ is the mean electron density today and we used $\bar{n}_e(\eta) = \bar{n}_e(0) a(\eta)^{-3}$.\\

With the Limber approximation \cite{Lim54}, or, equivalently using an approximate identity related to the completeness of spherical Bessel functions 
\begin{equation}
\int dk k^2 F(k) j_l(k \eta) j_l(k \eta') \approx \frac{\pi}{2} \da^{-2} \delta_D(\eta -\eta') F(k)|_{k=l/\da} \, ,
\end{equation}
the projected SZ power spectrum can be written as
\begin{equation}
C_{l}^{\sz}= \int dr \, \frac{W^{\sz}(r)^2}{d_A^2(r)}P_{\rm \Pi \Pi}\left(k=\frac{l}{d_A(r)};z(r)\right) \, ,
\end{equation}
where $d_A$ is the comoving angular-diameter distance, which in the case of a flat-cosmological model is $d_A = \eta$.
We have defined the SZ weight function
\begin{equation}
W^{\sz}(\eta) \equiv g_{\nu} \frac{\sigma_T}{m_e} \frac{k_B \bar{n}_e (0) }{a(\eta)^2 }\, .
\end{equation} 
The Limber approximation assumes
that $F(k)$ is a slowly varying function when compared to oscillations in $j_l(x)$. Under this assumption, contributions to the power spectrum arise only from correlations on equal time slices of spacetime.\\

Given that the three-dimensional power spectrum of pressure can be written as a combination of 1- and 2-halo terms and the fact that the 1-halo
term dominates the fluctuations, we can simplify the integral further and write it as \cite{KomSel02b}
\begin{equation}
C_l^{\sz}=\int dz \frac{d^2V}{dzd\Omega} \int dM \int dc \frac{d^2n}{dMdc} \Pi_l^2(z) \, ,
\end{equation}
where $d^2V/dzd\Omega$  is the physical volume per unit redshift per unit solid angle and $\Pi_l$ is related to the three-dimensional Fourier transform of the  electron pressure profile via 
\begin{equation}
\Pi_l(z) = \frac{W^{\sz}(z) \Pi_e\left(k=\frac{l}{d_A},z\right)}{d_A^2} \, .
\end{equation}
Using the projected angular size of a cluster $\theta=r/d_A$, one can simplify further and reduce $\Pi_l(z)$ to the following two-dimensional integral 
\begin{equation}
\Pi_l(z) = 2 \pi W^{\sz}(z)  \int_0^{\theta_{\rm out}} \theta d\theta J_0\left[\left(l+\frac{1}{2}\right) \theta \right] 
\int_{-r_{\rm out}}^{+r_{\rm out}} \Pi(r_\parallel,r_\perp=d_A \theta)dr_\parallel \, ,
\end{equation}
where we have rewritten the volume integral Eq.~\ref{equ:pressFT} in terms of two separate integrals involving the line of sight distance through
an individual halo, $r_\parallel$, and the 2-dimensional projected extent of the cluster, $r_\perp=d_A \theta$.
Here, $r_{\rm out}$ is the outer radius of the halo,
which with a pressure profile that falls off with distance can be taken as $r_{\rm out} \rightarrow \infty$, while $\theta_{\rm out}=r_{\rm out}/d_A$.\\

We now describe the ingredients required for the halo-based calculation of the SZ power spectrum in detail.

\begin{figure}[t]
\centerline{\psfig{file=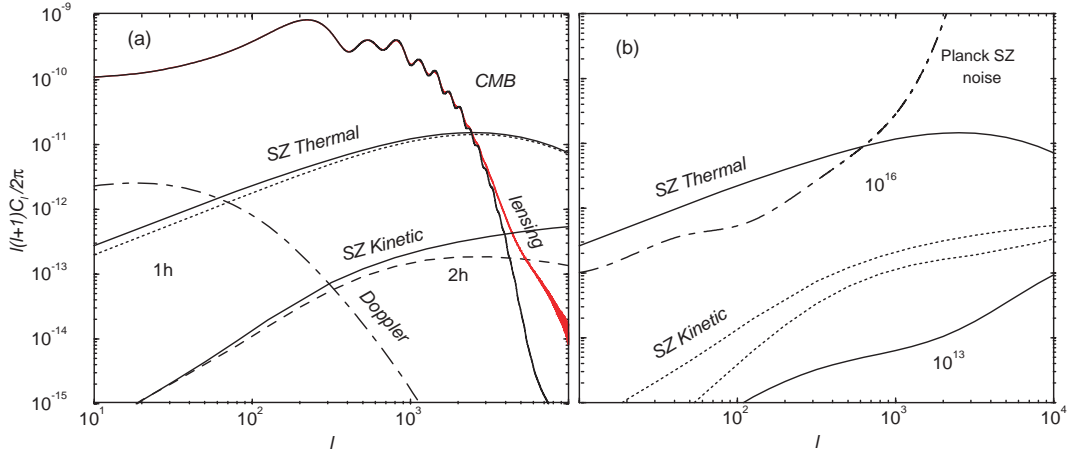,width=6in}}
\caption{The angular power spectra associated with the thermal and kinetic SZ
effects. As shown in (a), the thermal
SZ effect is dominated by individual halos, and thus, by the single
halo term (1h, dotted line), while the kinetic effect is dominated by the large scale
structure correlations depicted by the 2-halo term (2h, dashed line). In (b), we show
the mass dependence of the thermal and kinetic SZ effects with a
maximum mass of $10^{16}$ and $10^{13}$ M$_{\sun}$. The thermal SZ
effect (solid lines) is strongly dependent on the maximum mass, while the kinetic effect (dotted lines) is not since it receives most of its contribution from large scale correlations.}
\label{fig:szpower}
\end{figure}

\subsection{Halo Mass Function}

The mass function of the halo distribution is derived from the Press-Schechter formalism \cite{PreSch74}.
Press-Schechter theory is based on the idea that fluctuations in the linear density field with $\delta > \delta_{\rm c}$ detach from the local Hubble
expansion of the universe and collapse to form non-linear structure.  The prediction for the fraction of the volume that has collapsed is
\begin{equation}
f_{\rm coll} (M(R), z) = \frac{2}{\sqrt{2 \pi} \sigma(R,z)} \int_{\delta_{\rm c}}^{\infty} d \delta \, e^{-\delta^2/2\sigma^2(R,z)}\, ,
\end{equation}
where $R$ is the radius over which the density field has been smoothed, which is related to the halo mass by
$M(R) = \rho_{m} \frac{4\pi}{3} R^3$ with $\rho_{\rm m}$ the co-moving matter density of the universe.
The number density of halos is then found to be
given by \cite{PreSch74,SheTor99}
\begin{equation}
\frac{dn(M,z)}{dM} = -\frac{\rho_{m}}{M} \frac{d f_{\rm coll}(M(R),z)}{dM} = \frac{\rho_{\rm m}}{M}f(\nu)\frac{d\nu}{dM} \, ,
\label{eqn:massfunc}
\end{equation}
where
\begin{equation}\label{equ:fnu}
f(\nu) \equiv \sqrt{\frac{2A^2a^{2}}{\pi}}[1 + (a\nu^2)^{-p}]e^{-\frac{a\nu^2}{2}} \, .
\label{eqn:fnu}
\end{equation}

Here
\begin{equation}
\nu(M;z) \equiv \frac{\delta_{\rm c}(z)}{\sigma(M;z)} \, ,
\end{equation}
and
\begin{equation}
\delta_{\rm c}(z) \simeq \frac{3(12\pi)^{\frac{2}{3}}}{20}\left[1-\frac{5}{936}{\rm ln}\left(1+\frac{1-\Omega_{\rm m}}{\Omega_{\rm m}(1+z)^3}\right)\right]
\end{equation} is the critical density required for spherical
collapse at a redshift $z$ in a flat $\Lambda$CDM model (see e.g. Ref. \cite{Henry97} and references therein
for this and related formulae).
The variance in the density field smoothed with a top-hat filter of radius
$R=({3M}/{4\pi \rho_{\rm m}})^{\frac{1}{3}}$ is
\begin{equation}
\sigma^2(M;z) = G^2(z)\int \frac{dk}{k} \frac{k^3 P(k)}{2\pi^2} |W(kR)|^2 \, ,
\end{equation}
where
\begin{equation}
W(x)=\frac{3}{x^3}[{\rm sin}(x)- x{\rm cos}(x)] \, ,
\end{equation}
$P(k)$ is the linear matter power spectrum, and
\begin{equation}\label{equ:growth}
G(z) =
\frac{H(z)\int_{z}^{\infty}dz^{\prime}(1+z^{\prime})[H(z^{\prime})]^{-3}}{H_0\int_{0}^{\infty}dz^{\prime}(1+z^{\prime})[H(z^{\prime})]^{-3}}
\,
\end{equation}
is the linear theory growth factor (often also denoted $D(z)$).
$A$, $p$, and $a$ in equation (\ref{equ:fnu}) are constants, with
the canonical Press-Schechter (PS) and Sheth-Tormen (ST) mass functions corresponding to the parameters $(p=0,a=1)$ and $(p=0.3,a=0.707)$,
 respectively.   The normalization $A$ is determined by requiring mass
conservation such that
\begin{equation}
\frac{1}{\rho_{\rm m}}\int_0^{\infty}dM M \frac{dn}{dM} = \int_0^{\infty}d\nu f(\nu)=1 \, .
\end{equation}

\subsection{Dark Matter Density Profile}
The  halo mass function has to be supplemented by the dark matter density profile.
We define the dimensionless variable $u \equiv {r}/{r_{\rm s}}$, where $r_{\rm s}$ is a characteristic scale radius. The dark matter
profile within each halo is then defined as an NFW profile \cite{Navetal96} 
\begin{equation}
\rho_{\rm d}(r) \equiv\rho_{\rm s}\varrho_{\rm d}(u)=\rho_{\rm s}u^{-1}(1+u)^{-2}.
\end{equation}
Within the context of the spherical collapse model, the outer extent of the cluster is taken to be
the virial radius
\begin{equation}
r_{\rm v}=\left[\frac{3M}{4\pi\rho_{\rm m}(z)\Delta(z)}\right]^{\frac{1}{3}} \, ,
\end{equation}
where $\rho_{\rm m}(z) \propto (1+z)^3$ is the mean background matter density of the universe at redshift $z$, and
\begin{equation}
\Delta(z) \simeq 18\pi^2\left[ 1+\frac{88}{215} \left(\frac{ 1 - \Omega_{\rm m}}{\Omega_{\rm m}(1+z)^3}\right)^{\frac{86}{95}} \right]
\end{equation}
 is the overdensity of the halo relative to the
background density \cite{Henry97}.
The ratio of the virial radius to the scale radius is called the concentration parameter $c \equiv {r_{\rm v}}/{r_{\rm s}}$.
Together, $c$ and $M$ completely determine the dark matter distribution of a given halo.

\subsection{Concentration-Mass Distribution}

Numerical simulations indicate that clusters of a given halo mass have a range of concentration parameters. To account for this distribution we describe the number density of halos of mass $M$ and concentration $c$ by the bivariate
distribution function,
\begin{equation}
\frac{d^2 n}{dMdc}(M,c;z) \equiv \frac{dn}{dM}(M;z){\cal P}(c|M;z) \, ,
\end{equation}
where ${\cal P}(c|M;z)$ is the probability of a halo having
concentration $c$ given that it has mass $M$.
Numerical simulations \cite{Buletal01} find an approximately log-normal
distribution
\begin{equation}
{\cal P}(c|M;z) \, dc =\frac{{\rm exp}\left({{-\frac{[{\rm log}c-{\rm log}{\bar{c}(M;z)}]^2}{2\sigma_{{\rm log}c}^2}}}\right)}{\sqrt{2\pi}\sigma_{{\rm log}c}}\frac{dc}{{\rm ln}(10) c} \, ,
\label{eqn:pc}
\end{equation}
where the mean concentration parameter $\bar{c}$ is related to the halo mass
via
\begin{equation}
\bar{c}(M;z)=\frac{c_0}{1+z}\left[ \frac{M}{M_{*}(z=0)} \right]^{-\alpha_{c}} \, .
\end{equation}
$M_*(z)$ is the mass scale at which $\nu(M_*;z)=1$ and $c_0$ and $\alpha_c$ are constants whose numerical values are typically chosen to be $c_0=9$ and $\alpha_c=0.13$.
It is in general possible that $\sigma_{{\rm log}c}=\sigma_{{\rm
log}c}(M;z)$, as suggested by the scaling arguments of Ref. \cite{Verde01} and indeed a mass dependence
has been noted in numerical simulations \cite{Buletal01}.  However, for simplicity we fix the width of the concentration distribution to be $\sigma_{{\rm log}c}=0.14$ independent of mass.  

\subsection{Gas Pressure Profile}
\label{subsec:gaspress}

The gas density profile in terms of the dimensionless parameter $u$ is $\rho_{\rm g}(r) \equiv \rho_0 \varrho_{\rm g}(u)$.
The isothermal assumption for the gas
distribution used in Refs. \cite{Coo00,Coo01}  was found to be inconsistent with
the properties of numerically simulated clusters \cite{KomSel02a}. We therefore
implement a polytropic equation of state and write the gas pressure as
\begin{equation}
P_{\rm g}(r) \equiv P_0p(u) = P_0[\varrho_{\rm g}(u)]^{\gamma} \, ,
\end{equation}
with $\varrho_{\rm g}(0) = p(0) = 1 $, so that $\rho_0$ and $P_0$ correspond to the central density and pressure.
The gas profile within each halo has to self-consistently satisfy hydrostatic equilibrium,
\begin{equation}
\frac{1}{\rho_{g}}\frac{dP_{g}}{dr}=-\frac{GM(r)}{r^2} \, ,
\label{eqn:hydro}
\end{equation}
where $M(r)=\int_{0}^{r}dr^{\prime} 4\pi r^{\prime 2} [\rho_{g}(r^{\prime}) + \rho_{\rm d}(r^{\prime} )]$ is the total mass of baryonic gas and dark matter.
Rewriting Eq.~(\ref{eqn:hydro}) in terms of the dimensionless variables we introduced above leads to the following second order differential equation
\begin{equation}
\frac{d}{du}\left[\gamma u^2 \varrho_{\rm g}^{\gamma-2}\frac{d\varrho_{\rm g}}{du}\right] = - \lambda u^2[\varrho_{\rm d}(u)+\beta \varrho_{\rm g}(u)] \, ,
\label{eqn:diffeq}
\end{equation}
where
\begin{equation}
\beta \equiv \frac{\rho_0}{\rho_{\rm s}}
\label{eqn:beta}
\end{equation}
and
\begin{equation}
\lambda \equiv \frac{4\pi G \rho_0 \rho_{\rm s} r_{\rm s}^2}{P_0}  \, .
\label{eqn:lambda}
\end{equation}

In general $\varrho_{\rm g}=\varrho_{\rm g}(u ; \lambda, \beta, \gamma, c, f_{\rm b})$. To fix $\varrho_{\rm g}(u;c,f_{\rm b})$ for a given halo of concentration
$c$ and cosmological baryon fraction $f_{\rm b}={\Omega_{\rm b}}/{\Omega_{\rm m}}$
we impose the boundary condition that the gas density profile traces
the dark matter density profile in the outer region of the halo. This ansatz from Ref.~\cite{KomSel02a} ensures that the
hydrodynamical properties of the cluster resemble those found in simulations of clusters which
include both gas and dark matter.
In order to implement this boundary condition and determine $\lambda(c;f_{\rm b})$, $\beta(c;f_{\rm b})$, and $\gamma(c;f_{\rm b})$, we minimize the functional
\begin{equation}
\Psi[\varrho_{\rm g}(u;\lambda, \beta, \gamma, c, f_{\rm b})] =
\int_{\frac{1}{2}c}^{2c}\frac{du}{u}\left[s_{\rm g}(u)-s_{\rm d}(u)\right]^2 \, ,
\label{eqn:bigpsi}
\end{equation}
subject to the normalization constraint
\begin{equation}
\frac{\rho_{\rm g}(r_{\rm v})}{\rho_{\rm d}(r_{\rm v})}=\frac{\Omega_{\rm b}}{\Omega_{\rm d}}=\beta
c(1+c)^2\varrho_{\rm g}(c) \, ,
\label{eqn:normbc}
\end{equation}
where $s(u)={d{\rm ln}\varrho}/{d{\rm ln}u}$ is the local logarithmic slope of the density profile $\varrho(u)$.
We find that this procedure matches that of Ref. \cite{KomSel02a} when the gas self-gravity is neglected.

In Fig.~\ref{fig:SB} (left panel),
we present a brief comparison of our model and the isothermal model with results from numerical simulations done as part of the
Santa Barbara cluster comparison project \cite{Freetal99}. As shown in Fig.~\ref{fig:SB}, the isothermal cluster overestimates the gas density and pressure
in the inner region and also shows a departure at the outer extent of the cluster.  Comparing our model with that of Ref.~\cite{KomSel02b} we see
that including self-gravitational effects raises the pressure by $\sim 20 \%$ outside the cluster core, which can increase the amplitude of the angular power spectrum of the SZ effect by as much as 25-30\%.

\section{Comparison with Current CMB Data}

In 2002, two significant results on CMB temperature fluctuations at arcminute angular scales appeared. The
Cosmic Background Imager (CBI; \cite{Masetal02}) and the Berkeley-Illinois-Maryland Array (BIMA; \cite{Dawetal02}) measured an excess of small-scale power beyond
the damping tail of the primary CMB.
Interpreting the CBI and BIMA results as low redshift SZ contributions to the temperature anisotropies
allows constraints to be placed on cosmological parameters.
The cosmological implications of the small-scale excess in CBI were first studied in Ref. \cite{Bonetal02}
using numerical simulations of the SZ effect associated with unresolved clusters.
A more detailed analysis of the small-scale CBI data, based on semi-analytic calculations,
was presented in Ref. \cite{KomSel02b},
where a constraint was placed on the normalization of the matter power spectrum $\sigma_8$.\\

To study the constraints on cosmological parameters from small-scale anisotropies, one generally
performs a likelihood analysis.  Here, one must take into account the window functions of band-power from CBI and BIMA.
In order to compare data to theoretical predictions, we calculate the relative $\chi^2$ between data and models
\begin{equation}
\chi^2 = \sum_{i \leq j} (\hat{\bp_i} - \bp_i) C_{ij}^{-1} (\hat{\bp_j} - \bp_j) \, ,
\end{equation}
where $\hat{\bp_i}$ is a band-power estimate from recent observations, $\bp_i$
is the model prediction, and $C_{ij}^{-1}$ is the inverse covariance matrix. As we will now discuss, due to
the highly non-linear nature of galaxy clustering, one must take non-Gaussian contributions to the covariance into account.

\begin{figure}[t]
\centerline{\psfig{file=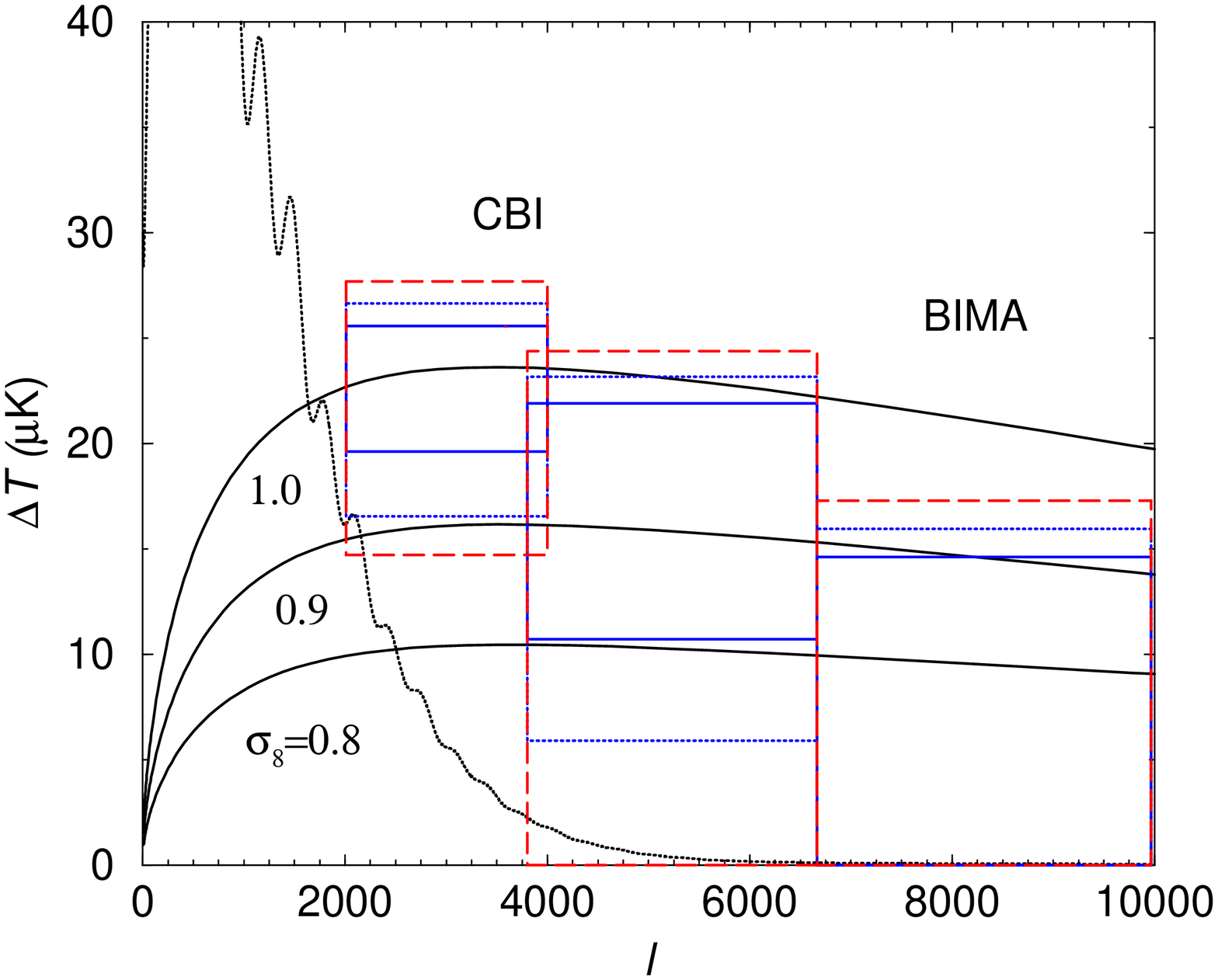,width=2.7in,angle=-90}
\psfig{file=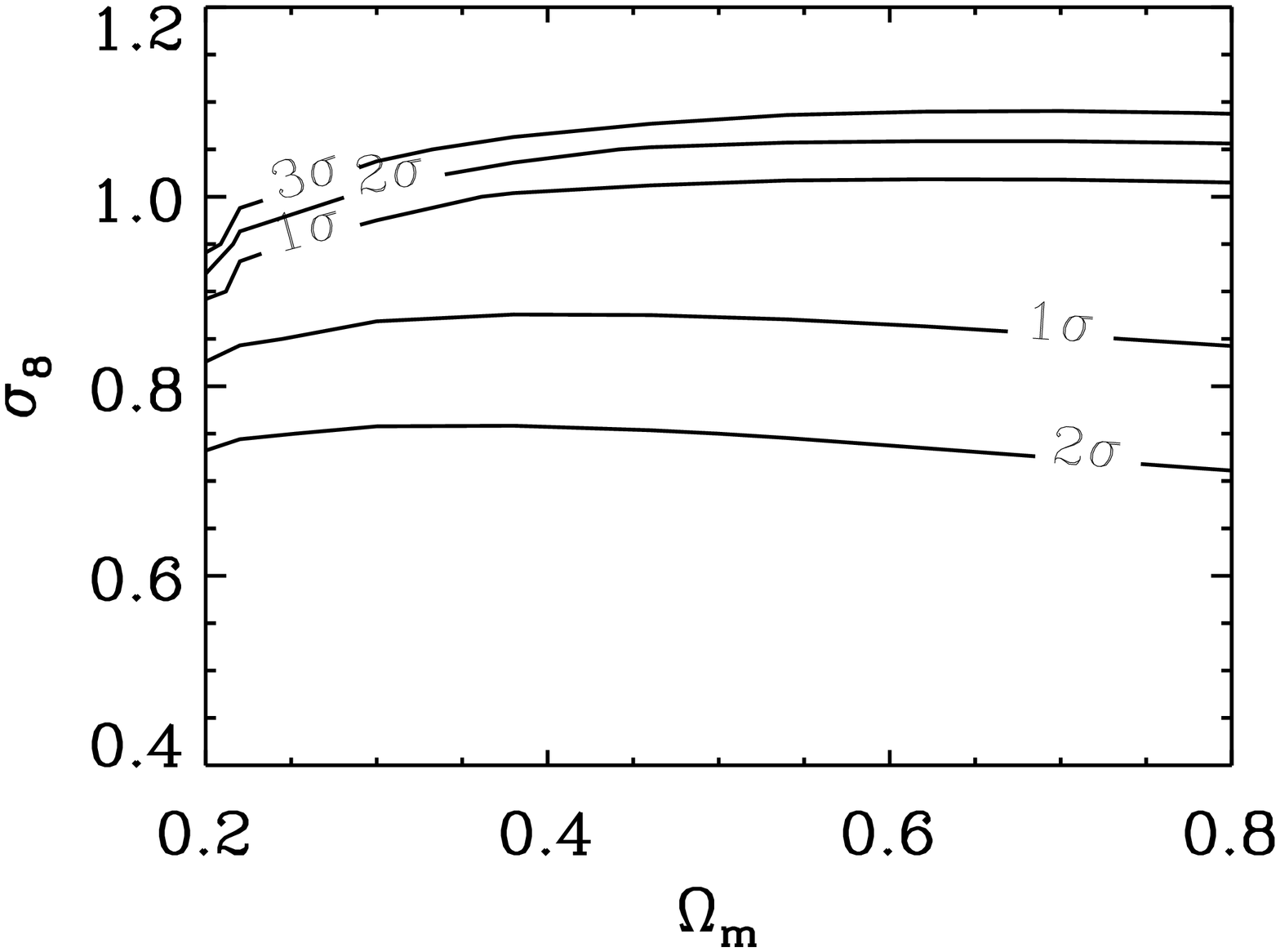,width=3.2in,angle=90}}
\caption{(a) Small-scale CMB anisotropies as observed by the CBI and BIMA experiments. The three
solid lines show predictions for the SZ effect for varying $\sigma_8$:  1.0 (upper), 0.9 (middle), and 0.8 (lower). For reference we show the predicted CMB primary anisotropies for our fiducial model. The solid error boxes show the published Gaussian errors, while the dotted
boxes include the non-Gaussian trispectrum.
When comparing data to predictions, we account for the full window
function of these observations. (b) Constraints on cosmological parameters. We consider constraints on $\Omega_{\rm m}$ and $\sigma_8$ using small-scale power and show
68\%, 95\%, and 99\% confidence limits.}
\label{fig:data}
\end{figure}

\subsection{Covariance}
\label{sec:covar}

The measurements of the power spectrum reported by the CBI and BIMA experiments are binned estimates of
power in multipole space with window functions $W(\vecl_i)$ at each band $i$, where $\vecl_i$ is a vector in a plane of the sky sufficiently small to be considered flat.  One can write these band-power estimates as 
\begin{equation}
\hat \bp_i = \int
{d^2 \vecl_i \over {A_{i}}} W(\vecl_i) W(-\vecl_i) \hat \Theta(\vecl_i) \hat \Theta(-\vecl_i) \, ,
\end{equation}
where $A_i = \int d^2 \vecl_i W^2(\vecl_i)$ is the window-weighted area of the two-dimensional shell in Fourier space
and $\hat \Theta(\vecl_i)$ is the Fourier transform of temperature fluctuations $\hat \Theta(\hat{\bf n}) \equiv \Delta T /T (\hat{\bf n})$ in 
the SZ map. 
Following Ref. \cite{CooHu01}, we can write the covariance matrix of these band-power measurements as 
\begin{equation}\label{eqn:variance}
C_{ij} = { (2\pi)^2 \over A} \left[{2 \bp_i^2 \over A_i}\delta_{ij}
+ {T^\sz_{ij} \over (2\pi)^2}\right]\, ,
\end{equation}
where
\begin{equation}
T^\sz_{ij}=
\int {d^2 \vecl_i \over A_{i}}
\int {d^2 \vecl_j \over A_{j}} W^2(\vecl_i) W^2(\vecl_j) T_\sz(\bfl_i,\bfl_j)\, 
\end{equation}
and $A$ is the total area of the survey in steradians. The first
term in equation (\ref{eqn:variance}) is the usual Gaussian sample variance and the
second term is the extra non-Gaussian contribution that arises from the
trispectrum associated with the SZ effect. The Gaussian noise also
includes an additional noise variance due to instrumental noise.
In the case of CBI and BIMA, the first term, including instrumental noise, is generally determined
during the band-power measurements. Thus, we only consider the extra covariance from the non-Gaussian
contribution. We again use the halo approach to model this term. At the small scales relevant for current data,
the one-halo term dominates \cite{Coo01}. We thus consider the trispectrum due to
single halos, 
\begin{equation}
T_\sz(\bfl_i,\bfl_j) \equiv T_\sz(\bfl_i,-\bfl_i,\bfl_j,-\bfl_j) = \int dr \frac{W^\sz(\rad)^3}{d_A^6(r)}T^{\rm 1h}_{\rm \Pi \Pi}\left(\frac{\bfl_i}{d_A(r)}, \frac{\bfl_j}{d_A(r)};z(r)\right) \, .
\end{equation}
Note that the projected trispectrum is generally a function of four $\bfl$ vectors and may therefore be represented by a general quadrilateral in
Fourier space. In the case of the power spectrum covariance we are only interested in terms involving $T_\sz(\bfl_i,-\bfl_i,\bfl_j,-\bfl_j)$,
i.e. parallelograms which are defined by either the length $l_{ij}=|\bfl_i-\bfl_j|$ or the angle
between $\bfl_i$ and $\bfl_j$. This can be noted from the fact that the correlator between
$\langle \hat \bp_i \hat \bp_j  \rangle$ involves the temperature fluctuations as
$\langle \hat \Theta(\vecl_i) \hat \Theta(-\vecl_i) \hat \Theta(\vecl_j) \hat \Theta(-\vecl_j) \rangle$
which is simply equal to the trispectrum with the above dependence.

Under the halo model the trispectrum involves four terms related to configurations with 1 to 4 halos.
On small scales the 1-halo term dominates, similar to the case of the power spectrum, and we can write this contribution to the
trispectrum responsible for power spectrum covariance as
\begin{eqnarray}
T^{\rm 1h}_{\rm \Pi \Pi}(\bfk_1,\bfk_2;z)= 
  \int  dM \int dc \,\frac{d^2n}{dMdc} \left|\Pi(|\bfk_1|, M, c ; z) \right|^2
\left|\Pi(|\bfk_2|, M, c ; z) \right|^2  \, .
\end{eqnarray}
In addition to the extra variance (when $|\bfl_i|=|\bfl_j|$), this non-Gaussian term also
contributes to the covariance and correlates band-power measurements between different
bins (when $|\bfl_i| \neq |\bfl_j|$). Note that in the case of the 1-halo term, the trispectrum is independent of the
angle between $\bfl_i$ and $\bfl_j$ and only depends on the amplitude of these two vectors, though, this is not the case in general
when other terms, such as 2-halo and higher, are involved.

In Fig.~4, we show the extra variance contribution  to the band-power estimates due to
non-Gaussianities. These non-Gaussian errors are such that they decrease with increasing sky area of the survey
and should be understood as arising from the rareness of galaxy clusters that contribute to the SZ anisotropies.
In Fig.~\ref{fig:data}, we show the angular power spectrum of the SZ effect along with the primordial anisotropies for our fiducial cosmological model.
Here we plot flat-band power per logarithmic interval in
$\mu$K such that
\begin{equation}
\Delta T = \sqrt{\left(\frac{l(l+1)}{2\pi}C_l\right)} T_{\rm CMB} \, ,
\end{equation}
where $T_{\rm CMB}=2.726 \times 10^6$ $\mu$K.
We also show flat-band power estimates of the
anisotropies at small angular scales from CBI \cite{Masetal02} and
BIMA \cite{Dawetal02}. The two error boxes represent the Gaussian error published by
both experimental groups and the total which includes the non-Gaussian error. For comparison, we show three predictions for the SZ power
spectrum for $\sigma_8$ ranging from 0.9 to 1.1. In general the SZ angular power spectrum $C_l^{\sz}$ scales approximately as $(\sigma_8)^7$ \cite{Seletal01}.

To extract cosmology, we consider a parameter set comprised only of
$(\Omega_{\rm m},\sigma_8)$, fixing all other cosmological parameters to fiducial values given by the WMAP best-fit \cite{Spergel:2003cb}.
The 1-$\sigma$ and 2-$\sigma$ constraints on this two-parameter space are shown in Fig.~\ref{fig:data}.
Our constraints are consistent with those obtained by Ref. \cite{KomSel02b} and suggested in
Ref. \cite{Bonetal02}. Marginalizing over values of $\Omega_{\rm m}$ in the range of 0.14 to 0.96,
we find $\sigma_8(\Omega_{\rm b} h/0.036)^{0.3}=0.95^{+0.10}_{-0.23}$ at 95\% confidence.
The data allow $\sigma_8$ values of around 1.0 to 0.85 at the 1$\sigma$ level, consistent with
complementary cosmological measurements. The inclusion of self-gravity and the distribution in concentration
increases power beyond the models in Ref. \cite{KomSel02a}.
Note that our results, while consistent with results from the analysis in Ref. \cite{Hol02a},
still allow a larger range of values for $\sigma_8$ than suggested there.\\

While small scale anisotropies can be interpreted as being due to the thermal SZ effect, this
interpretation is not unique. Arcminute-scale anisotropies can easily be generated at the
surface of last scattering if the primordial power spectrum had a feature. It is, however, important to keep in mind that scale-dependent features in the primordial power spectrum may be unnatural in specific models of inflation. Other effects producing small-scale anisotropies
include primordial voids at last scattering. If the reionization optical depth is high, as
suggested by the recent WMAP results \cite{kogut}, and the reionization process is associated with massive stars,
then a substantial small-scale signal could also be generated by the thermal SZ effect associated with the first supernova
bubbles \cite{Oh:2003sa}.  Similarly, at low redshifts, the secondary signal could be due to
foreground sources such as unresolved point sources. 

As discussed in Ref.~\cite{CooMel02}, to uniquely
determine if the contribution is associated with the thermal SZ effect, one needs to establish the frequency spectrum with
multi-frequency observations, or consider cross-correlation studies with tracers of large scale
structure. If the contribution is generated at last scattering, then one does not expect
a significant correlation with, say, the low redshift galaxy distribution, while a
strong correlation is expected, if arcminute scale fluctuations are related to the thermal SZ effect.
A similar cross-correlation analysis would be required to separate the thermal SZ effect in galaxy clusters from
the effect related to the first supernovae. The latter signal would not be resolved, while with high
resolution SZ maps, the removal of resolved SZ clusters will lead to a substantial decrease in the amplitude
of the SZ power spectrum, by which the presence of SZ supernovae can be explored.

In addition to the power spectrum, due to the highly non-linear nature of the thermal SZ effect, a significant non-Gaussianity is
induced by SZ contributions to arcminute-scale CMB fluctuations \cite{CooHu00,Coo00,Coo01} which can be measured from
the bispectrum, the three-point correlation function in Fourier space, or more easily, using
measurements of the skewness, among others. The SZ effect is also strongly correlated with angular deflections associated with
gravitational lensing modifications of CMB anisotropies \cite{CooHu00,SpeGol99,Cooray:2001ps}.
Future experiments such as Planck will be able to measure some of these
higher-order effects.

\section{Kinetic SZ Contribution to the CMB Power Spectrum}

We next provide a brief discussion of the kinetic SZ contribution to the CMB power spectrum.
The kinetic SZ effect results from the modulation of the large-scale velocity field by non-linear density structures such as galaxy clusters. 
In Fourier space, the effect is described as a convolution of the density and the velocity fluctuations (cf. Eq.~(\ref{equ:kSZ})). If we consider the flat-sky limit in which the Limber approximation \cite{Lim54} applies, then we can write the associated power spectrum as
\begin{equation}\label{equ:56}
C_l^\ksz = \frac{1}{8\pi^2} \int d\eta \, \frac{(g \dot{G}G)^2}{d_A^2} \,
I_v\left(k=\frac{l}{d_A}\right) \, .
\end{equation}
Recall that $g(\eta)$ and $G(\eta)$ are the visibility function (\ref{equ:visibility}) and the linear growth function (\ref{equ:growth}), respectively.
The mode-coupling integral is given by
\begin{equation}\label{equ:57}
I_v(k) = \int dk_1 \int_{-1}^{+1} d\mu \, \frac{(1-\mu^2)(1-2\mu
y_1)}{y_2^2} P_{\delta \delta}(k y_1) P_{gg}(k y_2) \, ,  
\end{equation}
where $P_{\delta \delta}$ and $P_{gg}$ denote the power spectra of perturbations in the dark matter and the cluster gas densities, 
respectively.  In the above, $\mu = \hat{\bf k} \cdot \hat{\bf k}_1$, $y_1 = k_1/k$ and $y_2 = k_2/k = \sqrt{1-2\mu y_1+y_1^2}$.
The wave vectors $\bfk_1$ and $\bfk_2 =\bfk-\bfk_1$ capture the convolution between velocity and gas density perturbations. 
We refer the reader to \cite{OstVis86,Coo01,DodJub95,JafKam98,Huetal95} for the details of the derivation of equations (\ref{equ:56}) and (\ref{equ:57}). 
 The velocity field power spectrum is related to the linear dark matter density field through the use of linear theory arguments involving the continuity
equation where ${\bf v} = -i \dot{G} \delta \bfk/k^2$, which was used in deriving Eq.~(\ref{equ:57}).

Since velocity fluctuations have a much larger coherence length than the non-linear gas density fluctuations traced by galaxy clusters, we consider the limit in which the velocity field is uncorrelated with the clusters.
Physically, this can thought of as the limit in which large-scale bulk flows are modulated by
small-scale overdensities.
In this small-scale limit, the dominant contributions to the integrals in equations (\ref{equ:56}) and (\ref{equ:57})   
arise from modes with $k_2 = |\veck - \veck_1| \sim k$ such that $y_1 \ll 1$ and $y_2 \rightarrow 1$,
and equation (\ref{equ:56}) reduces to
\begin{equation}
C_l^\ksz = \frac{1}{3}
\int d\eta \frac{(g \dot{G} G)^2}{d_A^2} P_{gg}(k) v_{\rm rms}^2 \, ,
\label{eqn:redflatsky}
\end{equation}
where $v_{\rm rms}^2$ is the rms of the (large scale) uniform bulk velocity
\begin{equation}
v_{\rm rms}^2 \equiv \int dk \frac{P_{\delta\delta}(k)}{2\pi^2} \, .
\end{equation}
The factor of $1/3$ in Eq. (\ref{eqn:redflatsky}) arises from the fact that the rms in each component is $1/3$
of the total velocity.\\ 

\begin{figure}[t]
\centerline{\psfig{file=cooray7.eps,width=4.2in}}
\caption{Line of sight projected maps of the thermal (left) and kinetic
(right) SZ effects (Figures reproduced from Ref.~\cite{Springel:2000bq}). The maps are $1^\circ \times 1^\circ$ and cover the same
field of view. Note that the thermal SZ map picks out massive halos, while
contributions to the kinetic SZ effect come from a wide range of masses. Unlike the
thermal SZ effect, which produces a negative decrement at Rayleigh-Jeans
wavelengths, the kinetic SZ effect varies between negative and positive
values depending on the direction of the velocity field.
Here, structures in red are moving towards the observer while those in
blue are moving away.}
\label{fig:tszksz}
\end{figure}

Just as in the halo model calculation of the thermal SZ effect, we split the power spectrum of perturbations in the cluster gas densities into 1-halo and 2-halo terms, $P_{gg} \equiv P_{gg}^{1h} + P_{gg}^{2h}$. For the thermal SZ effect we found that the 1-halo term was the dominant contribution and that correlations between halos could be neglected. For the kinetic SZ effect, however, we find that the 2-halo correlation term is most important and the 1-halo term can be neglected.
This is illustrated in Fig.~\ref{fig:szpower}, where we show our prediction for the CMB power spectrum associated with the kinetic SZ
effect and a comparison with the thermal SZ contribution.
It is observed that the kinetic SZ contribution is roughly an order of magnitude
smaller than the thermal SZ contribution. We also see that the thermal SZ effect is dominated by individual halos, while the kinetic SZ effect is dominated by large scale structure correlations depicted by the 2-halo term. We understand these results as follows:
The temperature anisotropies associated with the thermal SZ effect are proportional to the temperature weighted electron density (or the electron pressure, cf. equation (1)) and are therefore dominated by the most massive clusters with a correspondingly high electron temperature.
The kinetic SZ contribution is independent of the cluster temperature and depends only on the electron density and the cluster velocity.
Contributions to the kinetic SZ effect therefore come from baryons
tracing all scales down to small mass halos and the CMB power spectrum associated with the kinetic SZ rescattering is dominated by the large scale correlations of the halos.

The different mass dependence
of the two effects suggests that wide-field thermal and
kinetic SZ maps will show characteristic differences in that
massive halos, or clusters, will be clearly visible in a map tracing the thermal SZ effect, while the large scale structure will be more evident in
a kinetic SZ map. As shown in the thermal and kinetic SZ maps
in Fig.~\ref{fig:tszksz} from \cite{Springel:2000bq}, numerical simulations are in fact consistent
with this picture (see also \cite{Silva:2000yr}).

In addition to the contribution due to the line of sight motion of halos,
there is an effect resulting from halo rotations as discussed by
\cite{Cooray:2001vy}. Recent studies have considered this non-uniform
motion of cluster gas with numerical simulations (e.g.
\cite{Nagai:2002nw}).
The signal is considerably smaller than the bulk flow kinetic SZ effect and therefore unlikely to be
of interest to current and next-generation measurements of
arcminute-scale CMB anisotropies.

\section{SZ Contributions to the Polarized CMB}

Polarization of CMB anisotropies is 
generated when the CMB photons scatter off free electrons. The
polarization therefore traces the ionization history of the
universe. The universe was fully ionized at early times, before last
scattering occurred at a redshift of about $z_{\rm rec} \approx 1100$,
after which the universe became neutral and radiation and matter
decoupled. Last scattering generated the {\it primary} polarization of the
microwave background radiation. Since the generation of polarization is a strictly
causal process this signal is expected to peak at the horizon scale at
recombination. Causality doesn't allow a polarization signal on larger
scales. Such a signal, however, has recently been measured by the
Wilkinson Microwave Anisotropy Probe (WMAP) \cite{kogut}. This
{\it large-scale} {\it secondary} polarization is interpreted as a
signature of a late time reionization of the universe at a redshift of
$z \sim 10-20$. Since reionization leads to free electrons in galaxy
clusters we also expect {\it small-scale} secondary polarization \cite{Hu99,BauCooKam}. 
In Ref.~\cite{CooBau}, we revisited the problem of measuring a
CMB-induced polarization signal towards {\it resolved} galaxy
clusters. This was previously studied by Refs. \cite{SunZel80,SazSun,Challinor,Audit}.

Linear polarization of the cosmic microwave background is generated
through rescattering of the temperature quadrupole.
In a cosmological model with dark energy the quadrupole evolves
between the last scattering surface ($z=1100$) and us ($z=0$) due the
integrated Sachs-Wolfe (ISW) effect. The quadrupole-induced
polarization signal therefore provides an opportunity to probe dark energy through the ISW
effect. Kamionkowski and Loeb~\cite{KamLoeb,Port} have considered the possibility of using multiple such 
measurements to reduce the cosmic variance  uncertainty in the CMB temperature quadrupole.
The connection between properties of dark energy, the
quadrupole at the cluster redshift and polarization in the direction
of galaxy clusters is summarized here.

\subsection{CMB-induced Polarization towards Clusters}
\label{sec:pol}

CMB polarization towards clusters is generated when the incident radiation has a
non-zero temperature quadrupole moment. The two dominant origins for this quadrupole moment
are: (a) a projection of the {\it primordial} CMB quadrupole to the cluster location, 
and (b) a local {\it kinematic} quadrupole from cluster peculiar motion.
Towards a sufficiently large sample of galaxy clusters, we can write the total
rms degree of polarization as $P_{\rm Total}^2 = P_{\rm Prim}^2 + P_{\rm Kin}^2$, Ref.~~\cite{SazSun}, 
where
\begin{equation}\label{e:prim}
P_{\rm Prim} \propto \langle \tau \rangle \,Q^{\rm rms}(z) \propto \sqrt{C_2(z)} \, ,
\end{equation}
\begin{equation}
P_{\rm Kin} \propto g(x)\, \langle \tau \rangle \,\langle\beta_t^2\rangle \, . 
\end{equation}
$\tau =\sigma_T\int d\ell n_e(\ell) $ is the scattering optical depth of a cluster with $\ell$ the physical line of
sight distance through the cluster and $n_e(\ell)$ the electron number density. 
Since we are averaging over large samples of clusters,
we consider the sample-averaged optical depth, $\langle \tau \rangle$.
$\beta_t=v_t/c$ gives the transverse component of the cluster velocity
and $g(x)=(x/2)\coth(x/2)$, with $x \equiv h\nu/k_B T_{\rm CMB}$, is the frequency 
dependence of the kinematic effect.
With the optical depth in individual clusters determined by complementary
methods, such as the Sunyaev-Zel'dovich (SZ, \cite{SunZel80}) effects,
one can invert the measured polarization, equation~(\ref{e:prim}), to
obtain the rms CMB-quadrupole, $Q^{\rm rms} (z) \equiv (5 C_2/4\pi)^{1/2}$, at the cluster redshift.

\begin{figure}[t]
\centerline{\psfig{file=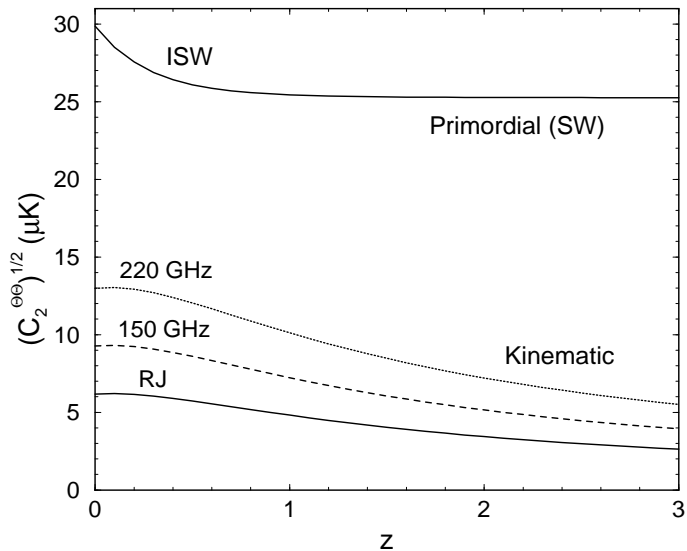,width=3.5in,angle=-90}}
\caption{The temperature quadrupole, $C_2^{\Theta \Theta}$, ($\Theta \equiv \Delta T/T$), as a function of
     redshift. We show both the primordial
     and the kinematic quadrupole. The bottom
     kinematic-quadrupole curve is for $g(x)=1$, as appropriate
     for the Rayleigh-Jeans (RJ) part of the frequency spectrum, and
     the dashed and dotted curves are, respectively, for
     frequencies 150 and 220 GHz.}
\label{fig:PowSpec}
\end{figure}

\begin{figure}[!t]
\centerline{\psfig{file=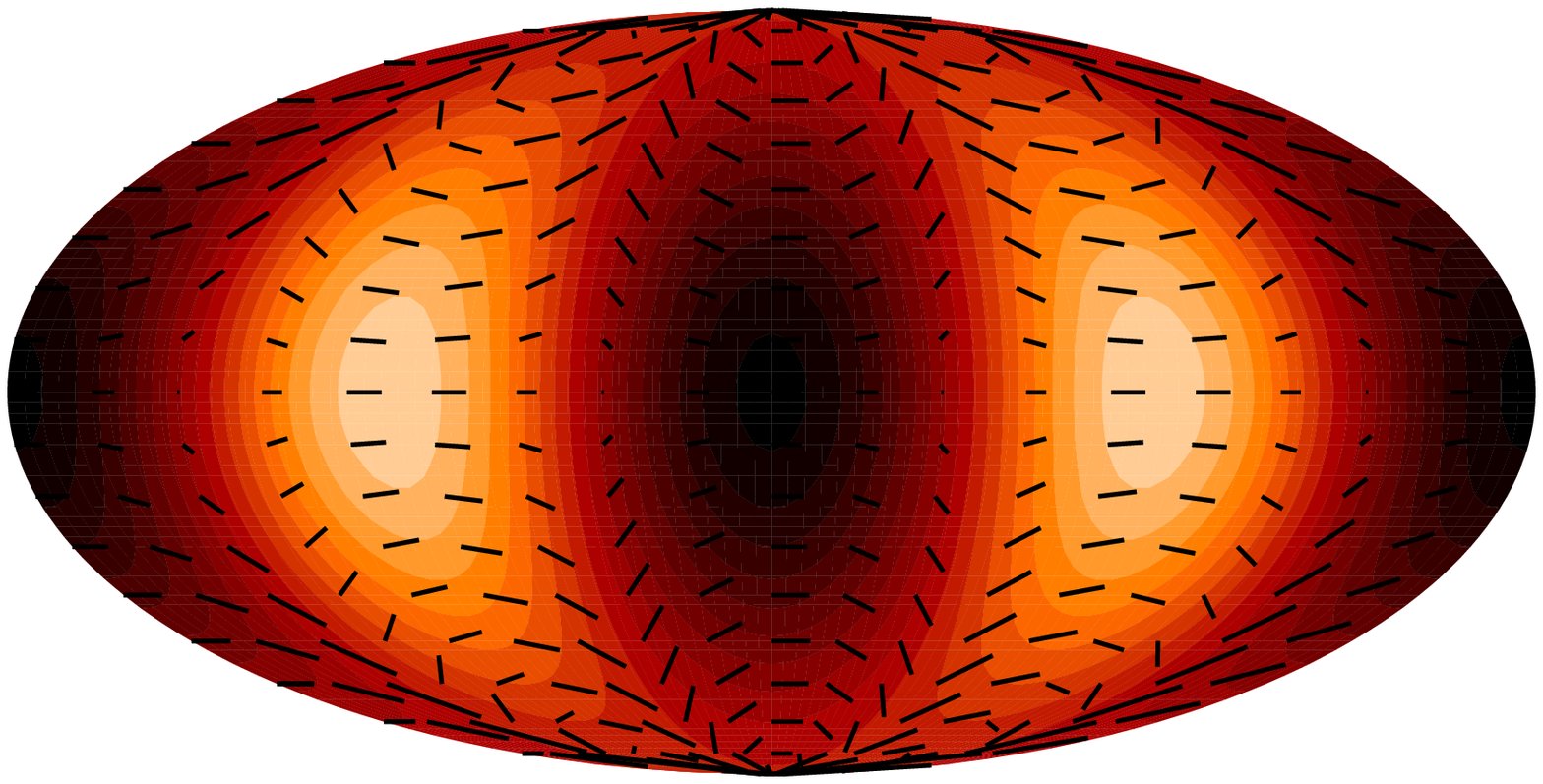,width=2.7in}\psfig{file=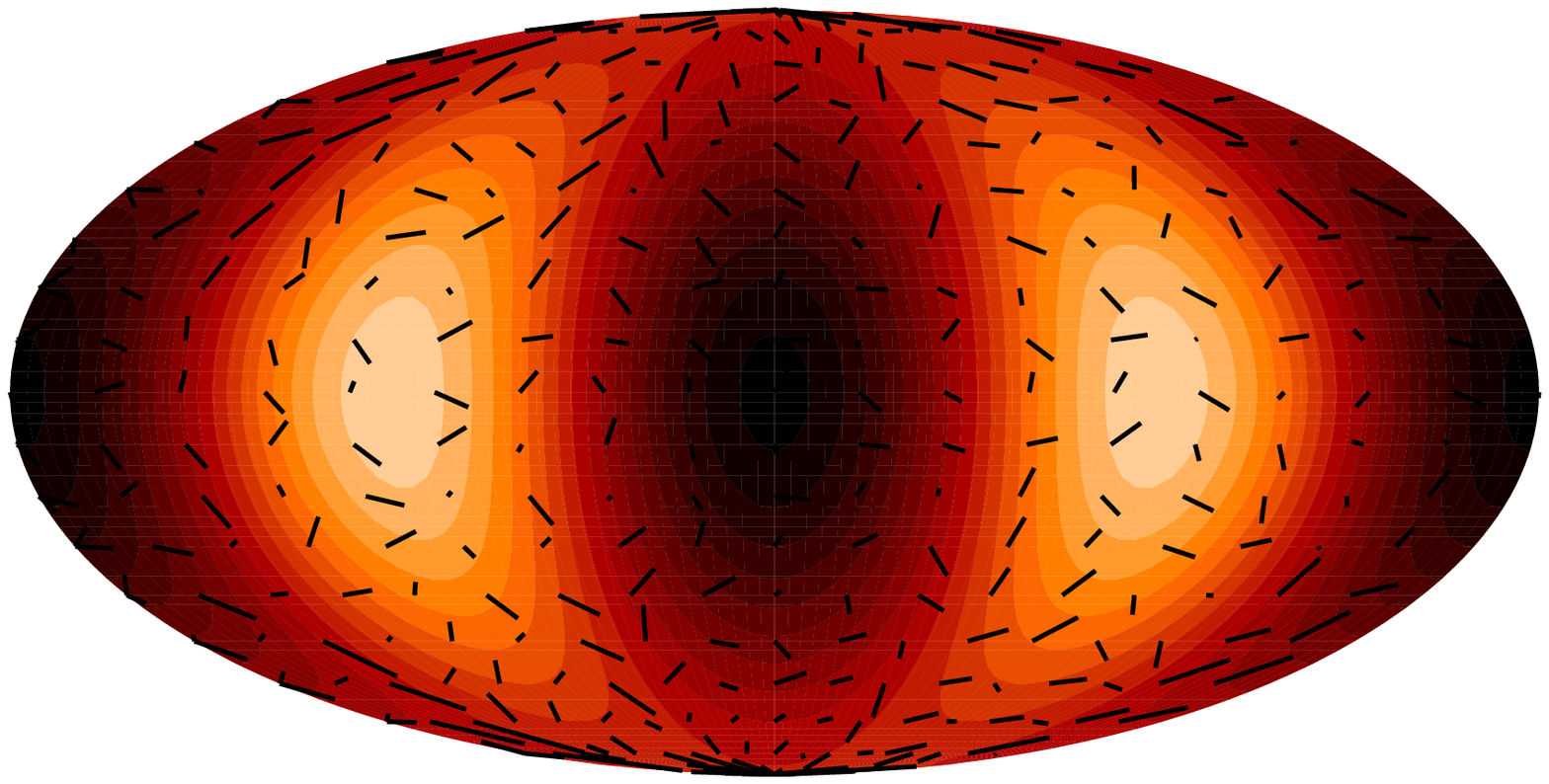,width=2.7in}}
\caption{CMB polarization due to galaxy clusters.
The polarization vectors give a representation of the expected
polarization from a cluster at the corresponding location in the sky.
The scale is such that the maximum length of a line corresponds to a
polarization of 4.9$\tau$ $\mu$K, where $\tau$ is the optical depth of the cluster.
The background color represents the temperature quadrupole. The left
map shows
the resulting polarization contribution due to scattering of the
primordial  temperature quadrupole alone.
The map for the total polarization contribution (primordial
and kinematic) is indistinguishable from this map.
To make the kinematic polarization visible,
we arbitrarily increase its amplitude by a factor of 100 (right map).
The primordial polarization
dominates the total contribution even at high frequencies where the
kinematic quadrupole is boosted due to its spectral dependence.}
\label{fig:prim}
\end{figure}

\subsection{Primordial Quadrupole}

The primordial CMB-quadrupole is dominated by two effects.
The Sachs-Wolfe (SW, \cite{SacWol67}) effect arises as a combination of gravitational
redshift and time-dilation effects and can be viewed as a direct
projection of the conditions at last scattering with no evolution
after that time:
\begin{equation}
\left(\frac{\Delta T}{T}\right)^{{\rm SW}} = {\Phi(\eta=\eta_{\rm ls}) \over 3}
\, ,
\end{equation}
where $\Phi$ is the Newtonian potential.
The integrated Sachs-Wolfe (ISW) contribution arises along the photon path from 
the time of last scattering to today, as
the CMB photons pass through a time-varying potential: 
\begin{equation}
\left(\frac{\Delta T}{T}\right)^{{\rm ISW}} =  2 \int_{\eta_{\rm ls}}^{\eta_0} \dot{\Phi} \, d\eta\, .
\end{equation}
Effectively, the photon receives a shift in energy because the potential it falls 
into is different from the potential it must climb out of.
The ISW effect is absent in a matter-dominated, critical-density
universe (Einstein-de Sitter). In a universe with dark energy ($w \equiv p/\rho < 0$) or a cosmological constant, $\Lambda$, ($w=-1$) 
the ISW effect leads to an increase in power on large scales.
The expected redshift evolution of the quadrupole, $C_{l=2}(z) = C_{l=2}^{{\rm SW}}(z)+C_{l=2}^{{\rm ISW}}(z)$, is hence characterized by a rise at 
low redshifts ($z<1$), the time at which the universe becomes dark energy dominated.

We calculate $C_2^{\Theta\Theta}(z)$, $\Theta \equiv \Delta T/T$, following Ref.~\cite{Hu99} for our fiducial
$\Lambda$CDM cosmology and show the result in Fig.~\ref{fig:PowSpec}.
Note that at redshift $z=0$, with the power-spectrum tilt fixed at unity,
COBE finds $C_2^{\Theta\Theta}(z=0)=(27.5 \pm 2.4 \,\mu{\rm K})^2$
\cite{Bennett:gg}. At higher redshifts, the mean quadrupole moment evolves due to the
integrated Sachs-Wolfe effect and possibly, if the power
spectrum is not flat, due to any scale dependence since the
quadrupole probes smaller distances at earlier times.  
The primordial quadrupole has a coherence length
comparable to the horizon.  We thus expect the polarization signal
measured on a ${\cal O}(10^\circ)$ patch of sky may differ by order unity from that on a different
${\cal O}(10^\circ)$ patch of sky. 
Finally, note that Thomson scattering from
cold electrons will not change the photon frequency. Thus,
there will be no frequency dependence of the rescattering of the primordial quadrupole.

\subsection{Kinematic Quadrupole}

The origin of the kinematic polarization effect is understood as follows. Consider electrons 
moving with peculiar velocity, $\beta=v/c$, relative to the rest frame defined by the CMB.
The Doppler-shifted spectral intensity of the CMB in the mean electron rest frame is 
\begin{equation}
I_{\nu} = C \frac{x^3}{e^{x \gamma(1+\beta \mu)}-1}\, ,
\end{equation}
where $x \equiv h \nu/ k_B T_{\rm CMB}$, $\gamma = (1-\beta^2)^{-1/2}$ and
$\mu$ is the cosine of the angle between
the cluster velocity and the direction of the incident CMB photon.
When expanded in terms of Legendre polynomials, the intensity distribution is
\begin{equation}
I_{\nu} = C \frac{x^3}{e^x-1} \left[ I_0 + I_1 \mu +
\frac{e^x(e^x+1)}{2(e^x-1)^2}x^2 \beta^2 \left(\mu^2-\frac{1}{3}\right) + \dots
\right] \, ,
\end{equation}
which contains the necessary quadrupole component under which scattering generates
polarization. 

Unlike the primordial quadrupole, the kinematic quadrupole has a frequency
dependence which we denote by $g(x)$, and using the expansion of the
intensity distribution, in
temperature units instead of intensity units, one can show
$g(x)=(x/2)\coth(x/2)$. 

The $a_{22}$ quadrupole moment related to the cluster transverse motion (in a coordinate system in
which $z$ is along the line of sight) is
\begin{equation}\label{equ:a22}
     a_{22} = g(x) \sqrt{4 \pi \over 30} v_t^2\, e^{2 i \phi_v},
\end{equation}
where $\phi_v$ is the orientation angle for $v_t$ on the plane
of the sky.  To obtain this result, note that in a coordinate
system in which the ${\bf \hat z}$ axis is aligned with the
cluster's motion, the quadrupole dependence of the radiation
temperature is $g(x) v_t^2 (\mu^2-1/3) = g(x) v_t^2 (2/3)
\sqrt{4 \pi/5} Y_{2,0}$,
where $\mu$ is the cosine of the angle between the radiation
direction and the ${\bf \hat z}$ direction.  However, in a
coordinate system in which the ${\bf \hat z}$ axis is taken to be along the line
of sight, $(\mu^2-1/3)=-(1/3)\sqrt{4 \pi/5} Y_{2,0}+ \sqrt{2 \pi /
15}(Y_{2,2}+Y_{2,-2})$.  Thus, the coefficient of $Y_{2,2}$, the
component of the quadrupole moment that gives rise to
polarization in our direction, is only a fraction $\sqrt{3/8}$
of the total quadrupole moment.

Fig.~\ref{fig:prim} illustrates the
primordial and kinematic quadrupole contributions using all-sky
maps of the expected polarization.
In these plots, each polarization vector should be considered as a representation of the polarization towards a
cluster at that location. The polarization pattern created by scattering of the
primordial quadrupole is uniform and traces the underlying temperature quadrupole distribution.
For the kinematic quadrupole we assume, for illustrative purposes,
a transverse velocity field with $\langle \beta_t \rangle \sim
10^{-3}$ corresponding to a velocity of 300 km sec$^{-1}$.
The polarization contribution due to
the kinematic quadrupole, however, is random due to the fact that
transverse velocities are uncorrelated\footnote{The
correlation length of the velocity field ($\sim$ 60 Mpc) correlates
velocities within regions of 1$^{\circ}$ when projected to a redshift of
order unity.}. 
This explains the randomization of the all-sky polarization map when a
significant kinematic contribution is included.
It should be noted, however, that the kinematic effect has been scaled by a factor of $100$ to make it visible in Fig.~\ref{fig:prim}.
Therefore, as shown in Fig.~\ref{fig:prim},  the primordial polarization  dominates the total contribution even at high frequencies where
the kinematic quadrupole is increased due to its spectral dependence.
Also, the spectral dependence of the kinematic quadrupole contribution,
$g(x)$, gives a potential method to separate the two polarization effects \cite{Dod97}. 
This is similar to component separation suggestions in
the literature as applied to temperature observations, such as the
separation of the thermal SZ-effect from dominant primordial
fluctuations \cite{Cooetal00}.

\begin{figure}[t]
\centerline{\psfig{file=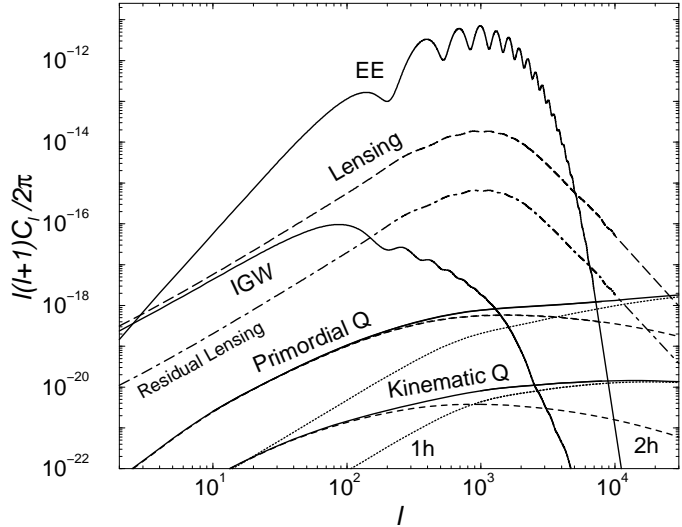,width=3.5in,angle=-90}}
\caption{
     Polarization power spectra due to the
     rescattering of the primordial and kinematic quadrupoles.
     We break total power in
     each of these cases into the 1- (1h; dotted lines) and 2-halo (2h; dashed lines) terms
     under the halo-based approach used here.  While at large
     angular scales correlations
     between halos dominate, at small angular scales of order few
     arcminutes and below, contributions are dominated by the
     1-halo term.  For comparison, we also show primordial
     polarization power spectra for E and B-modes involving
     dominant scalar (E-mode) and tensor (B-mode) contributions
     respectively. The tensor contribution to B-modes due to
     inflationary gravitational waves (IGW) assumes an energy scale
     for inflation of $10^{16}$ GeV. The long-dashed curve is the
contribution to B-modes of polarization resulting from the cosmic shear conversion of
power in E-modes, while
the dot-dashed line labeled ``Residual Lensing'' represents the noise
contribution after optimally
subtracting the lensing contribution using
higher-order statistics.}
\label{fig:cl}
\end{figure}

\subsection{CMB-induced Cluster Polarization and Dark Energy}

Secondary CMB polarization in the direction of distant galaxy cluster may provide an opportunity of study the evolution of growth in the late, dark energy dominated universe \cite{CooBau}, \cite{Baumann:2003xb}, \cite{CooHutBau}. This will constrain the time-evolution of the dark energy equation of state and therefore might contribute to a better understanding of the properties and the nature of dark energy.\\

The basic idea is to use galaxy clusters as tracers of the local temperature quadrupole and statistically detect its rms value. Imagine a future experiment measuring the polarization towards resolved clusters. 
For each cluster we also measure its redshift and the optical depth through the 
thermal SZ effect. We do this for a large sample of clusters, bin the resulting 
data into redshift intervals and average over all sky. If the bin size can be 
chosen small enough this will allow us to reconstruct the rms temperature quadrupole 
as a function of redshift. Multi-frequency observations will allow us to separate 
the primordial quadrupole from the contaminant kinematic contribution with only 
a factor of $\sim 2$ enhancement in the instrumental noise \cite{CooBau}.

For the best-fit $\Lambda$CDM model with $70\%$ dark energy, the quadrupole leads to a maximum
primordial polarization of $P_{{\rm Prim}} \sim 4.9\tau$ $\mu$K. Since
the kinematic polarization scales as $P_{{\rm Kin}} = 0.27 g(x)
(\beta_t/0.001)^2 \tau$ $\mu$K, we expect the CMB-induced signal to be
dominant (as illustrated in Fig.~\ref{fig:prim}). Factors that could make the primordial 
and the kinematic polarization more comparable are the frequency boost 
of the kinematic effect, a more optimistic estimate of $\langle \beta_t\rangle$
and a low value of the CMB-quadrupole. However, even if the signals were of comparable magnitude the random
orientations and characteristic frequency dependence of the kinematic
polarization would allow a reliable extraction of the primordial CMB-quadrupole.

\begin{figure}[!ht]
\centerline{
\includegraphics[height=3.0in, width=1.5in, angle=-90]{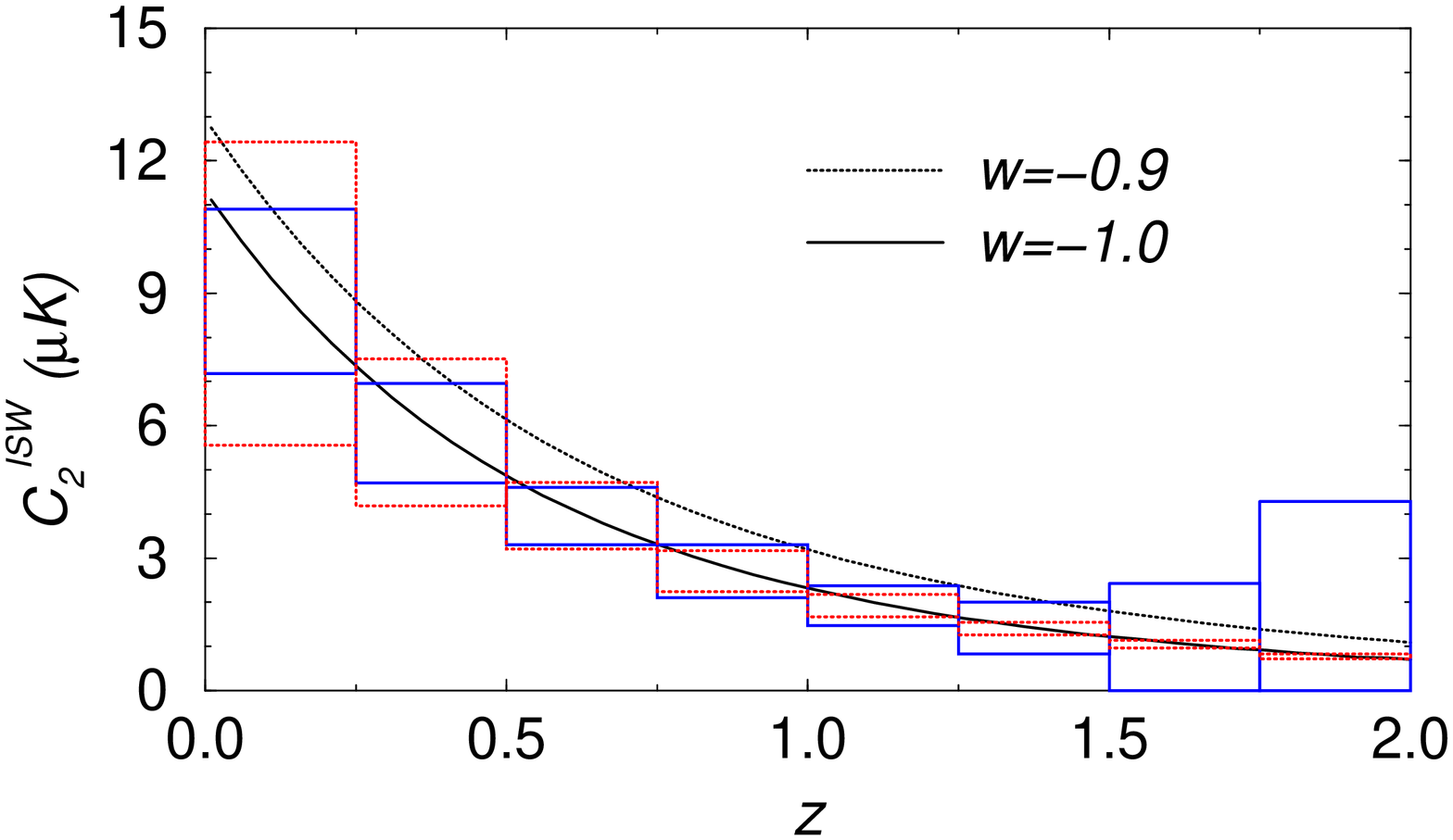}
\\[0.20cm]}
\centerline{
\includegraphics[height=1.8in, width=1.5in, angle=-90]{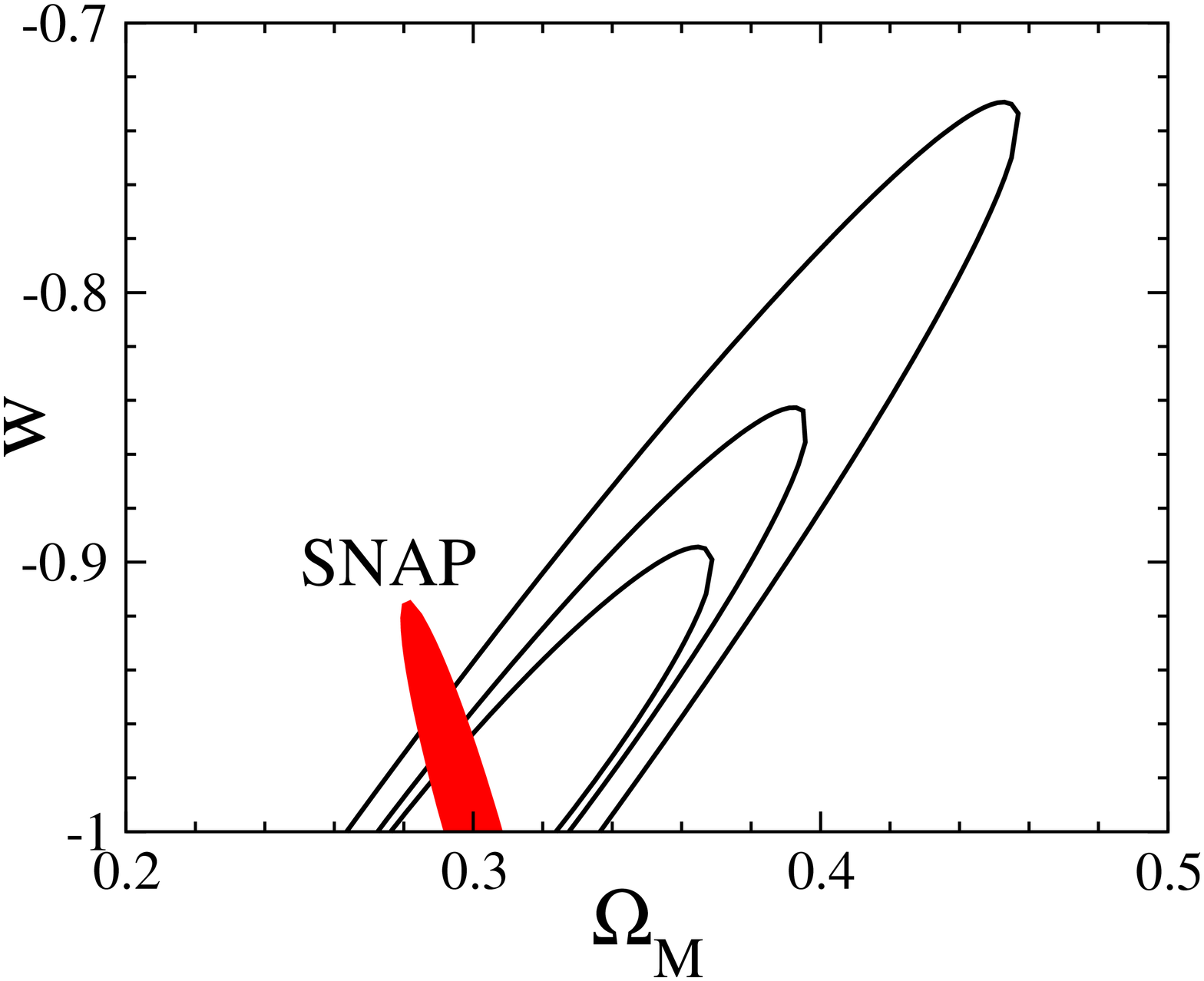}\nobreak
	\hspace{-0.4cm}
\includegraphics[height=1.8in, width=1.5in, angle=-90]{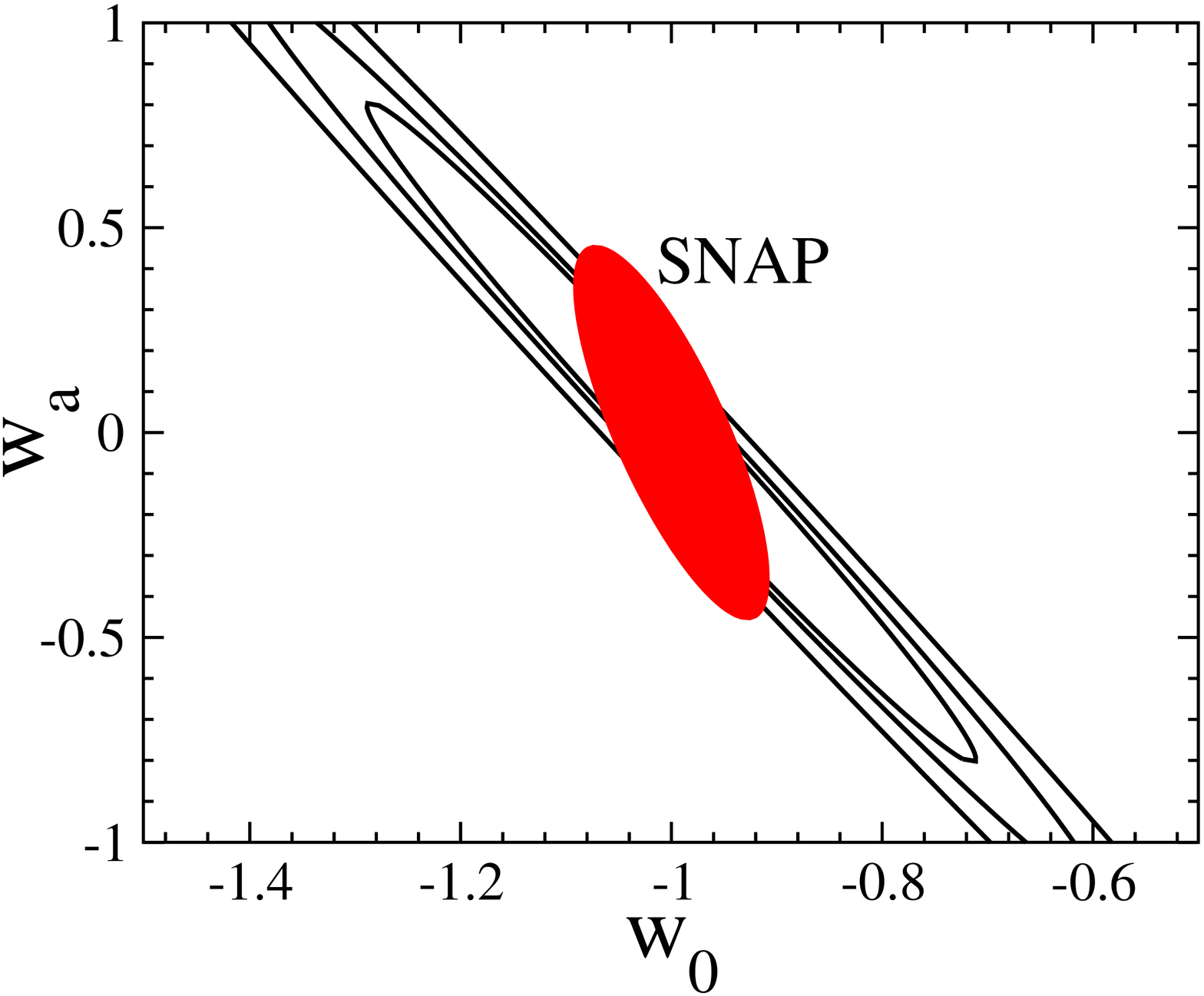}
}
\caption{{\it Top panel}: Projected ISW quadrupole as a function of redshift. 
The solid error bars assume a reconstruction with clusters down to
10$^{14}$ M$_{\sun}$ in an area of 10$^4$ deg$^2$ with an instrumental
noise of 0.1 $\mu$K. The dotted lines show the cosmic-variance for an
all-sky reconstruction computed from the number of independent volumes
sampled by clusters at each redshift bin~\cite{KamLoeb}.  {\it Bottom
left}: Parameter errors from the projected ISW quadrupole measurements,
assuming $w={\rm const}$. The small ellipse is for the case shown in
top panel (with cosmic variance added in quadrature), while the two
larger ellipses assume a factor of 3 and 10 increase in the
instrumental noise contribution, respectively.  For comparison, we
also show the constraints expected from SNAP. When the most optimistic 
polarization information is added, SNAP's constraints on $w$ improve by a factor of 3.
{\it Bottom right}: Same as bottom left, but
for $w_0$ and $w_a$ of the parameterization $w(z)=w_0+w_az/(1+z)$ and assuming an additional prior
$\sigma(\Omega_{\rm M})=0.01$.}
\label{fig:isw}
\end{figure}

In \cite{CooBau}, we assessed the potential detectability of the ISW
signature with cluster polarization, while
in \cite{CooHutBau}, the measurement of cosmological parameters related to the dark energy
using cluster polarization data from, say, the planned CMBpol mission, was discussed. 
The measurement of dark energy properties is aided by the fact that
one probes the redshift evolution of the ISW contribution through the growth rate of the gravitational potential.
The latter provides the most sensitive probe of dark energy when
compared to all other cosmological probes considered so far \cite{CooHutBau}. This
comes from the fact that the growth rate of the gravitational potential is directly proportional to
the dark energy equation of state, while quantities such as the
distance or the growth factor involve, at least, one integral of this
quantity.\\

Establishing the ISW effect and reconstructing its redshift evolution poses a significant experimental challenge.
The advent of polarization sensitive bolometers, however, suggests
that a reliable reconstruction of the ISW signature is within reach
over the next decade. Given the strong dependence of the ISW effect on the background
cosmology, cluster polarization can eventually be used as a powerful probe of
the dark energy. A detailed analysis of the potential of this method
for extracting information on the dark energy equation of state, $w(z)$, is presented in \cite{CooHutBau} and illustrated in Figure~\ref{fig:isw}.

\subsection{Power Spectra of SZ Polarization}

In addition to statistical studies to reconstruct the primordial quadrupole from CMB polarization measurement in the direction of resolved galaxy clusters, in the limit that clusters are unresolved,
power spectrum measurements would provide an approach to study the polarization signal.
In general, a polarization map will consist of a measurement of the Stokes parameters $Q(\bn)$ and
$U(\bn)$ as functions of position $\bn$ on some patch of the
sky.  We can construct Fourier components $Q({\bf l})$ and
$U({\bf l})$ by
\begin{equation}
     X({\bf l}) = \int d^2 \bn~e^{i {\bf l} \cdot \bn } X(\bn) \, ,
\end{equation}
where $X \equiv Q, U$, and ${\bf l}$ is a vector in the plane of a region of the
 sky
sufficiently small to be considered flat.  Since the Stokes
parameters $Q$ and $U$ depend on the choice of axes, we consider
the rotationally invariant combinations,
\begin{eqnarray}
     E({\bf l}) &=& \cos(2\phi_{\mathbf l}) Q({\bf l})+\sin(2\phi_{\mathbf l})
     U({\bf l}) \nonumber \\
     B({\bf l}) &=& \cos(2\phi_{\mathbf l}) U({\bf l})-\sin(2\phi_{\mathbf l})
     Q({\bf l}) \, ,
\end{eqnarray}
where $\phi_{\mathbf l}$ is the angle between ${\bf l}$ and the chosen x-axis
in the plane of the sky.  We define the angular power spectra
$C_l^{EE}$ and $C_l^{BB}$ from
\begin{equation}
     \langle Y(\vecl) Y(\vecl') \rangle = (2\pi)^2
     \delta_D(\vecl - \vecl') C_l^{YY} \, ,
\end{equation}
where $Y\equiv E,B$, and the angle brackets denote an average
over all realizations of the density field.

If the radiation quadrupole moment is smooth over the region of sky we are considering, then
\begin{eqnarray}\label{equ:EE}
     \langle E (\vecl) E(\vecl') \rangle &=& \langle
     B(\vecl)B(\vecl') \rangle  \nonumber \\
     &=& \frac{1}{2}  
     \left( \langle Q(\vecl) Q(\vecl') \rangle + \langle
     U(\vecl)U(\vecl') \rangle \right) \, ,
\end{eqnarray}
while
\begin{equation}\label{equ:EB}
     \langle E(\vecl)B(\vecl') \rangle = \frac{1}{2} \left(
     \langle Q(\vecl)U(\vecl') \rangle  -  \langle
     Q(\vecl)U(\vecl') \rangle \right) = 0 \, ,
\end{equation}
where the latter equality is consistent with parity
conservation.  These results can be derived by noting that if
the quadrupole moment is constant, then the orientation of the
polarization is constant.  If so, we may choose our axes on the
sky so that $U=0$.  Then, $E(\vecl) \propto \cos(2 \phi_{\mathbf l})$, and
$B(\vecl) \propto \sin(2 \phi_{\mathbf l})$, but when averaged over
the orientation angle $\phi_{\mathbf l}$, we recover Eq. (\ref{equ:EE}).

We now proceed to calculate the power spectra induced by
reionization.  The polarization in direction $\bn$ due to
scattering from free electrons is an integral along the line of
sight \cite{Kosowsky:1994cy},
\begin{equation}
     Q(\hat n) - i U(\hat n) = \sqrt{3 \over 40 \pi}
     \int \, d\eta {d \tau(\eta\bn,\eta) \over d\eta} a_{22}(\eta)\, ,
\label{eqn:projection}
\end{equation}
where $\eta$ is the comoving distance, $(d\tau/d\eta)(\eta\bn,\eta) =
\sigma_T n_e(\eta\bn,\eta) a(\eta)$, $a(\eta)$ is the scale factor
at a comoving distance $\eta$, $n_e(\eta \bn,\eta)$ is the
free-electron density at direction $\bn$ at distance $\eta$, and
$\sigma_T$ is the Thomson cross section. 

In Eq.~(\ref{eqn:projection}), 
$a_{22}(\eta)$ is the radiation quadrupole moment at distance $\eta$.  More
precisely, $a_{22}(\eta)$ is the coefficient of the
spherical harmonic $Y_{22}(\theta,\phi)$ in a spherical-harmonic
expansion of the radiation intensity in a coordinate system in
which the line of sight is the ${\bf \hat z}$ direction.  Note
that we take $a_{22}(\eta)$ to be a function of distance only, and
not direction, consistent with our assumption that the
quadrupole is coherent over a large patch of the sky.  
Since we use Limber's approximation below, in which angular correlations are
induced only by spatial separations at the same distance, the
variation of $a_{22}(\eta)$ with distance can be included
consistently.

With the polarization written as a projection along the line of sight, Eq.~(\ref{eqn:projection}), the angular
power spectrum follows in the flat-sky limit from Limber's
equation \cite{Lim54},
\begin{eqnarray}
     C_l^{EE} &=& C_l^{BB} \nonumber \\
     &=& \frac{3}{80 \pi} 
     \int_0^{z_{\rm rei}} dz \, \frac{d^2V}{d\Omega dz} |a_{22}(z)|^2\,
     P^{(t)}_{\tau\tau} \left(\frac{l}{d_A},z\right) \, ,
\label{eqn:cl}
\end{eqnarray}
where $P^{(t)}_{\tau\tau}$ is the power spectrum of $d\tau/d\eta$,
proportional to the power spectrum of the electron density
$n_e$, and the integral is taken up to the redshift $z_{\rm rei}$ of reionization using
the comoving differential volume given by $d^2V/d\Omega dz$.

In Fig.~\ref{fig:cl}, we summarize results related to the polarization power spectra from galaxy
clusters. In the case of the primordial quadrupole, we replace
$|a_{22}(z)|^2$ in Eq.~(\ref{eqn:cl}) by its
expectation value, $C_2^{\Theta\Theta}(z)$, the variance of the temperature
quadrupole at redshift $z$ (shown in Fig.~\ref{fig:PowSpec}).
In the case of the kinematic quadrupole, we replace $|a_{22}(z)|^2$ by
the expectation value of equation (\ref{equ:a22}),
\begin{equation}
    \VEV{|a_{22}|^2} = {4 \pi \over 30} g^2(x)\VEV{v_t^4} = {16 \pi
     \over 135} g^2(x) v_{\rm rms}^4\, ,
\label{eqn:vkin}
\end{equation}
where we have used
$\VEV{v_t^4}=\VEV{(v_x^2+v_y^2)^2}=(8/9) v_{\rm rms}^4$, since
$\VEV{v_x^2}=\VEV{v_y^2}=v_{\rm rms}^2/3$,
$\VEV{v_x^4}=\VEV{v_y^4}=3\VEV{v_x^2}^2=v_{\rm rms}^2/3$, and
$\VEV{v_x^2 v_y^2}=\VEV{v_x^2} \VEV{v_y^2}=v_{\rm
rms}^4/9$.  We calculate the linear-theory rms peculiar velocity by integrating over the power
spectrum, as in the case of the SZ kinetic temperature anisotropy calculation.
According the halo-clustering
model \cite{CooShe02}, peculiar-velocity fields are correlated
over large distances and the non-linear corrections to $v_{\rm rms}^2$ are small. The resulting linear-theory rms kinematic
quadrupole is also shown in Fig.~\ref{fig:PowSpec} assuming $g(x)=1$ as relevant for Rayleigh-Jeans (RJ) part of the
frequency spectrum when $x \rightarrow 0$, and for $\nu=150$ GHz
and $\nu=220$ GHz.

In Fig.~\ref{fig:cl}, note that the secondary polarization discussed here contributes
equally to E- and B-modes (cf. equation (\ref{eqn:cl})).  While the 1-halo term dominates at arcminute angular
scales and below, correlations between halos are important
and determine the total effect due to secondary
polarization at angular scales corresponding to a few
degrees.   The dependence of our results on the inclusion of the 2-halo
term is consistent with the result obtained for the temperature
power spectra from the kinetic SZ effect \cite{Coo01}, while it is inconsistent with the thermal
SZ effect, where contributions are dominated by the 1-halo term
over the whole range of angular scales. The latter behavior is
explained by the fact that the thermal SZ effect is highly dependent
on the most massive halos, while the kinetic SZ effect, and the
secondary polarization signals calculated here, are independent of the gas temperature and thus can depend on
halos with a wider mass range.

As shown in Fig.~\ref{fig:cl}, the secondary E-mode polarization
is several orders of magnitude below the E polarization from the
surface of last
scattering.  The secondary polarization is therefore
unlikely to be a source of confusion when interpreting
polarization contributions to E-modes.  The amplitude of the
primary effect in B-modes, due to gravitational waves, is highly
uncertain and depends on the energy scale of inflation \cite{KamKos}. For
illustration, we show in Fig.~\ref{fig:cl} the inflationary
gravitational wave (IGW) signal assuming an energy scale for
inflation of $E_{\rm infl}=10^{16}$ GeV, where the amplitude of the
power spectrum scales as $E_{\rm infl}^4$.
At large angular scales the secondary polarization is several
orders of magnitude below the peak of this hypothetical IGW
polarization signal.  If the energy scale of inflation is
lowered considerably, say to $E_{\rm infl} < 10^{15}$
GeV, then we might guess that the secondary polarization could
ultimately constitute a background.

As also shown in
Fig. \ref{fig:cl}, however, there is a contribution to the B-mode power
spectrum that arises from conversion of the primary E-modes to B-modes by gravitational lensing \cite{SeljakZaldarriaga}, and
this is considerably larger than the secondary polarization from galaxy clusters.
Moreover, we also show (the dot-dash curve) the contribution to
the irreducible B-mode power spectrum that remains even after
the lensing has been optimally subtracted with higher-order
correlations \cite{Kesdenetal,Knox,CooKes,Hu:2001fa}.  This residual lensing
power spectrum is considerably larger than the polarization from
reionization.  Thus, the secondary SZ polarization effects are
unlikely to be a factor for either gravitational-lensing or
gravitational-wave studies with B-modes.

\acknowledgments

AC thanks the organizers of the summer school, F. Melchiorri and Y. Rephaeli, for an
enjoyable and productive conference and for the invitation to present the research discussed in this review.  We would like to thank Marc Kamionkowski and Dragan Huterer for important collaboration upon which parts of this review are based.


\begin{thebibliography}{99}
\frenchspacing

\bibitem{SunZel80}
R.~A.~Sunyaev and Ya.~B.~Zel'dovich, MNRAS {\bf
190}, 413 (1980).

\bibitem{Freedman:2000cf}
W.~L.~Freedman {\it et al.},
Astrophys.\ J.\  {\bf 553}, 47 (2001).

\bibitem{Spergel:2003cb}
D.~N.~Spergel {\it et al.},
preprint [arXiv:astro-ph/0302209].

\bibitem{Caretal96}
        J. E. Carlstrom, M. Joy and L. Grego, \ApJ, {\bf 456}, L75 (1996).

\bibitem{Jonetal93}
M. Jones, R. Saunders, P. Alexander et al., Nature, {\bf 365}, 320 (1993).

\bibitem{Kai84}
        N. Kaiser, \ApJ, {\bf 282}, 374 (1984).

\bibitem{CooHu00}
A.~R.~Cooray and W.~Hu,
[arXiv:astro-ph/9910397].


\bibitem{OstVis86}
        J. P. Ostriker and E. T. Vishniac, ApJ, 306, L51 (1986);
        E. T. Vishniac, \ApJ, {\bf 322}, 597 (1987).


\bibitem{Aghanim:1996ib}
N.~Aghanim, F.~X.~Desert, J.~L.~Puget and R.~Gispert,
[arXiv:astro-ph/9604083].

\bibitem{Gruzinov:1998un}
A.~Gruzinov and W.~Hu,
[arXiv:astro-ph/9803188].

\bibitem{Santos:2003jb}
M.~G.~Santos, A.~Cooray, Z.~Haiman, L.~Knox and C.~P.~Ma,
Astrophys.\ J.\  {\bf 598}, 756 (2003)
[arXiv:astro-ph/0305471].

\bibitem{ColeKai88}
S.~Cole and N.~Kaiser, Mon.\ Not.\ Roy.\ Astron.\ Soc.\ {\bf 233}, 637 (1988);
E.~Komatsu and T.~Kitayama,  Astrophys.\ J.\ {\bf 526}, L1 (1999).

\bibitem{Coo00}
A.~Cooray, \PRD\ {\bf 62}, 103506 (2000).

\bibitem{Coo01}
A.~Cooray, \PRD\ {\bf 64}, 043516 (2001).

\bibitem{KomSel02b}
E.~Komatsu and U.~Seljak,  \MNRAS\ {\bf 336} 1256 (2002).

\bibitem{CooShe02}
A.~Cooray and R.~Sheth,
Phys.\ Rept.\  {\bf 372}, 1 (2002)
[arXiv:astro-ph/0206508].


\bibitem{KomSel02a}
  E.~Komatsu and U.~Seljak, \MNRAS\ {\bf 327} 1353 (2002).

\bibitem{White02}
        M.~White, L.~Hernquist, and V.~Springel, preprint [arXiv:astro-ph/0205437].

\bibitem{EisHu99} D.~Eisenstein and W.~Hu, \ApJ\ {\bf 511}, 5  (1999).

\bibitem{Moetal97}
        H.~J.~Mo, Y.~P.~Jing, and S.~D.~M.~White, \MNRAS\ {\bf 284}, 189 (1997).

\bibitem{Lim54}  D.~Limber, \ApJ\ {\bf 119}, 655 (1954).

\bibitem{PreSch74}
        W.~H.~Press and P.~Schechter, \ApJ\ {\bf 187}, 425 (1974);

\bibitem{SheTor99}
        R.~K.~Sheth and G.~Tormen, \MNRAS\ {\bf 308}, 119 (1999).

\bibitem{Henry97}
        J.~P.~Henry, \ApJ\ {\bf 534}, 565 (2000).

\bibitem{Navetal96}
        J.~Navarro, C.~S.~Frenk, and S.~D.~M.~White, \ApJ\ {\bf 462}, 563 (1996);
        B.~Moore, T.~Quinn, F.~Governato, J.~Stadel, and G.~Lake,  \MNRAS\ {\bf 310}, 1147 (1999).

\bibitem{Buletal01}
J.~S.~Bullock, T.~S.~Kolatt, Y.~Sigad, R.~S.~Somerville, A.~V.~Kravtsov, A.~A.~Klypin, J.~R.~Primack, A.~Dekel, \MNRAS\ {\bf 321}, 559 (2001); R\
.~H.~Wechsler, J.~S.~Bullock, J.~R.~Primack, A.~V.~Kravtsov, A.~Dekel, \ApJ\ {\bf 568}, 52 (2002); Y.~P.~Jing and Y.~Suto, \ApJL\ {\bf 529}, 69 \
(2000).

\bibitem{Verde01} L.~Verde, M.~Kamionkowski, J.~J.~Mohr, and A.~J.~Benson, \MNRAS\ {\bf 321}, L7 (2001).

\bibitem{Freetal99} C.~S.~Frenk et al., \ApJ\ {\bf 525}, 554 (1999).

\bibitem{Masetal02}
  B.~S.~Mason et al., preprint [arXiv:astro-ph/0205384].

\bibitem{Dawetal02}
        K.~S.~Dawson et al., preprint [arXiv:astro-ph/0206012].

\bibitem{Bonetal02} J.~R.~Bond et al., preprint [arXiv:astro-ph/0205386].

\bibitem{CooHu01}
A.~Cooray and W.~Hu,
Astrophys.\ J.\  {\bf 554}, 56 (2001)
[arXiv:astro-ph/0012087].


\bibitem{Seletal01} U.~Seljak, J.~Burwell, and U.~-L.~Pen, \PRD\ {\bf 63}, 063001 (2001).

\bibitem{Hol02a}
G.~P.~Holder,
Astrophys.\ J.\  {\bf 578}, L1 (2002)
[arXiv:astro-ph/0207633].

\bibitem{Oh:2003sa}
S.~P.~Oh, A.~Cooray and M.~Kamionkowski,
Mon.\ Not.\ Roy.\ Astron.\ Soc.\  {\bf 342}, L20 (2003)
[arXiv:astro-ph/0303007].

\bibitem{CooMel02}
Phys.\ Rev.\ D {\bf 66}, 083001 (2002)
[arXiv:astro-ph/0204250].


\bibitem{SpeGol99}
        D. N. Spergel and D. M. Goldberg, \PRD {\bf 59}, 103001 (1999);
        D. M. Goldberg and D. N. Spergel, \PRD, {\bf 59}, 103002 (1999);
        M. Zaldarriaga and U. Seljak, \PRD, {\bf 59}, 123507 (1999);
        H. V. Peiris and D. N. Spergel, \ApJ, {\bf 540}, 605 (2000).


\bibitem{Cooray:2001ps}
A.~Cooray,
Phys.\ Rev.\ D {\bf 64}, 043516 (2001)
[arXiv:astro-ph/0105415].


\bibitem{DodJub95}
        Dodelson, S. Jubas, J. M. 1995, ApJ, {\bf 439}, 503.

\bibitem{JafKam98}
Jaffe, A. H., Kamionkowski, M. 1998, Phys. Rev. D., {\bf 58}, 043001.


\bibitem{Huetal95}
        Hu, W., Scott, D., Sugiyama, N., White, M. 1995, Phys. Rev. D., {\bf 52}, 5498.


\bibitem{Springel:2000bq}
V.~Springel, M.~J.~White and L.~Hernquist,
[arXiv:astro-ph/0008133].

\bibitem{Silva:2000yr}
A.~C.~d.~Silva, D.~Barbosa, A.~R.~Liddle and P.~A.~Thomas,
Mon.\ Not.\ Roy.\ Astron.\ Soc.\  {\bf 326}, 155 (2001)
[arXiv:astro-ph/0011187].

\bibitem{Cooray:2001vy}
A.~Cooray and X.~l.~Chen,
Astrophys.\ J.\  {\bf 573}, 43 (2002)
[arXiv:astro-ph/0107544].

\bibitem{Nagai:2002nw}
D.~Nagai, A.~V.~Kravtsov and A.~Kosowsky,
Astrophys.\ J.\  {\bf 587}, 524 (2003)
[arXiv:astro-ph/0208308].


\bibitem{kogut}
A.~Kogut {\it et al.},
preprint [arXiv:astro-ph/0302213].

\bibitem{Hu99}
W.~Hu,
Astrophys.\ J.\  {\bf 529}, 12 (2000).


\bibitem{BauCooKam}
D.~Baumann, A.~Cooray and M.~Kamionkowski,
New Astron.\  {\bf 8}, 565 (2003)
[arXiv:astro-ph/0208511].


\bibitem{CooBau}
A.~Cooray and D.~Baumann,
Phys.\ Rev.\ D {\bf 67}, 063505 (2003)
[arXiv:astro-ph/0211095].

\bibitem{Baumann:2003xb}
D.~Baumann and A.~Cooray,
New Astron.\ Rev.\  {\bf 47}, 839 (2003)
[arXiv:astro-ph/0304416].

\bibitem{SazSun}
S.~Y.~Sazonov and R.~A.~Sunyaev, MNRAS {\bf 310}, 765 (1999).

\bibitem{Challinor}
A.~Challinor, M.~Ford, and A.~Lasenby, MNRAS  {\bf 312}, 159 (2000).

\bibitem{Audit}
E.~Audit and J.~F.~L.~Simmons, MNRAS {\bf 305}, L27
(1999).

\bibitem{KamLoeb}
M.~Kamionkowski and A.~Loeb,
Phys.\ Rev.\ D {\bf 56}, 4511 (1997);
N.~Seto and M.~Sasaki,
Phys.\ Rev.\ D {\bf 62}, 123004 (2000)
[arXiv:astro-ph/0009222];

\bibitem{Port}
J.~Portsmouth,
Phys.\ Rev.\ D {\bf 70}, 063504 (2004) 
[arXiv:astro-ph/0402173].

\bibitem{SacWol67}
        R.~K.~Sachs and A.~M.~Wolfe, Astrophys.\ J.\  {\bf 147}, 73 (1967).

\bibitem{Bennett:gg}
C.~L.~Bennett et al.,
Astrophys.\ J.\  {\bf 436}, 423 (1994).


\bibitem{Dod97}
 S.~Dodelson, Astrophys.\ J.\  {\bf 482}, 577 (1997).

\bibitem{Cooetal00}
A.~Cooray, W.~Hu, and M.~Tegmark, Astrophys.\ J.\ {\bf 540}, 1 (2000).


\bibitem{CooHutBau}
A.~Cooray, D.~Huterer, and D.~Baumann,
preprint [arXiv:astro-ph/0304268].


\bibitem{Kosowsky:1994cy}
A.~Kosowsky, Ann. Phys.\  {\bf 246}, 49 (1996).

\bibitem{KamKos}
See, for example, M. Kamionkowski and A. Kosowsky,
Annu. Rev. Nucl. Part. Sci. {\bf 49}, 77 (1999), and references
therein.


\bibitem{SeljakZaldarriaga}
M.~Zaldarriaga and U.~Seljak,
Phys.\ Rev.\ D {\bf 58}, 023003 (1998).


\bibitem{Kesdenetal}
M.~Kesden, A.~Cooray and M.~Kamionkowski,
Phys.\ Rev.\ Lett.\  {\bf 89}, 011304 (2002)
[arXiv:astro-ph/0202434].


\bibitem{Knox}
L.~Knox and Y.~S.~Song,
Phys.\ Rev.\ Lett.\  {\bf 89}, 011303 (2002).

\bibitem{CooKes}
A.~Cooray and M.~Kesden,
New Astron.\  {\bf 8}, 231 (2003)
[arXiv:astro-ph/0204068];
M.~Kesden, A.~Cooray and M.~Kamionkowski,
Phys.\ Rev.\ D {\bf 67}, 123507 (2003)
[arXiv:astro-ph/0302536].

\bibitem{Hu:2001fa}
U. Seljak and M. Zaldarriaga, Phys. Rev. Lett. {\bf 82}, 2636
(1999); W.~Hu, Phys.\ Rev.\ D {\bf 64}, 083005 (2001);
W.~Hu and T.~Okamoto,
[arXiv:astro-ph/0111606];
C.~M.~Hirata and U.~Seljak,
Phys.\ Rev.\ D {\bf 67}, 043001 (2003)
[arXiv:astro-ph/0209489].

\end{thebibliography}
\end{document}